\documentclass[preprint,10pt]{article}

\usepackage{blindtext}
\usepackage[a4paper, left=1in, right=1.in, top=1.2in, bottom=1.2in]{geometry}
\usepackage{booktabs}
\usepackage{url}
\usepackage{subcaption}
\usepackage{listings}
\usepackage[ruled, linesnumbered]{algorithm2e}
\usepackage{enumitem}
\usepackage{xcolor}
\usepackage{float}
\usepackage{arydshln}
\usepackage{tikz}
\usepackage{eurosym}
\usetikzlibrary{positioning}
\usepackage{gensymb}
\usepackage{alphalph}
\usepackage{amsmath, amssymb, graphicx}
\usepackage{graphicx} 
\usepackage{hyperref}
\usepackage{authblk}

\newtheorem{example}{Example}[section] 
\newtheorem{definition}{Definition}[section]  
\newtheorem{theorem}{Theorem}[section]
\newtheorem{proof}{Proof}[section]

\newcommand{\card}[1]{\vert #1 \vert} 

\setlist[itemize]{align=parleft,left=0pt..1em}
\setlist[enumerate]{align=parleft,left=0pt..2em}

\newdimen\nodeDist
\title{RED2Hunt: an Actionable Framework for Cleaning Operational Databases with Surrogate Keys }

\author{
    \textbf{Mathilde Marcy, Jean-Marc Petit, Vasile-Marian Scuturici}\thanks{INSA Lyon, CNRS, UCBL, LIRIS, UMR5205, Villeurbanne, F-69621, France} \\
    \textbf{Jocelyn Bonjour}\thanks{INSA Lyon, CNRS, CETHIL, UMR5008, Villeurbanne, F-69621, France} \\
    \textbf{Camille Fertel, Gérald Cavalier}\thanks{CEMAFROID, Fresnes, F-94266, France}
}

\date{}

\begin{document}

\maketitle

\begin{abstract}
Surrogate keys are now extensively utilized by database designers to implement keys in SQL tables. 
They are straightforward, easy to understand, and enable efficient access, despite lacking any real-world semantic meaning.
In this context, complex redundancy issues might emerge and often go unnoticed as long as they do not affect the operational applications built on top of the databases. These issues become evident when organizations seek to leverage data science, posing significant challenges to the implementation of analytical projects.\\

This paper, grounded in real-world applications, defines the concept of artificial unicity and proposes \textit{RED2Hunt (RElational Databases REDundancy Hunting)}, a human-in-the-loop framework for identifying hidden redundancy and, if problems occur, cleaning relational databases implemented with surrogate keys. 
We first define the central and intricate notion of artificial unicity and then the RED2Hunt framework to address it. 
We rely on simple abstractions easy to visualize based on the so-called \emph{redun\-dancy profile} associated to some relations and the notion of \emph{attribute stability}. Quite interestingly, those profiles can be computed very efficiently in quasi-linear time. We have devised different metrics to guide the domain expert and an actionable framework to generate new redundancy-free databases. The proposed framework was implemented on top of PostgreSQL.
From the publicly available IMDB database, we have generated synthetic databases, implementing different redundancy scenarios, on which we tested RED2Hunt to study its scalability.
RED2Hunt has also been tested on operational databases implemented with surrogate keys. Lessons learned from these real-life applications are discussed.

\end{abstract}


\section{Introduction}

Data science’s popularity is spreading to all strata of the corporate sphere, having until recently been a privilege reserved to large organizations with specialized resources and expertise. Today, most small and medium enterprises (SMEs) are interested in valuating their data through business intelligence (BI) or artificial intelligence (AI). Despite this strong interest, data quality remains a major obstacle to the spread of data analytics within these companies \cite{ehrlinger2022survey}, as the exploitation of bad data carries a strategic and financial risk \cite{mit2017dataquality}. Redundancy in particular is a very common and significant issue found in operational data, causing underlying data quality issues such as inconsistency and inaccuracy.

Most data owned by those companies is stored into relational databases management systems (RDBMS) used to support the digitalization of their operations. 
The design of these databases' schemas is often entrusted to the IT professionals in charge of developing the operational applications, who commonly favor performance over analytical constraints. 
For instance, IT developers tend to make intensive use of surrogate keys which are easy to understand and enable efficient access but do not hold any meaning in real-life. Surrogate keys might lead to data quality issues.

\begin{example}
Let us consider the fictional  \textit{Perfect Pet} operational database schema described in Figure \ref{fig:dbschema}.
All keys declared in the data dictionary turn out to be surrogate primary keys (suffixed with SK). Consequently, the two foreign keys (suffixed by SFK and given as the left-hand side of inclusion dependencies whose right hand side is a key) are also surrogates. 

\begin{figure}[bt]
\tt{
\footnotesize
\begin{flushleft}
Microchip(\underline{id\_microchip (SK)}, number, implant\_date) \\
Animal(\underline{id\_animal (SK)}, species, breed, name, id\_microchip (SFK), gender, dob, weight, food, hash\_id) \\
Appointment(\underline{id\_appointment (SK)}, id\_animal (SFK), date, time, main\_reason) \\
\end{flushleft}

FKs:\\
Animal[id\_microchip] $\subseteq $ Michochip[\underline{id\_microchip}]\\
Appointment[id\_animal] $\subseteq $ Animal[\underline{id\_animal}]
}
\caption{Running DB example: \textit{Perfect Pet}'s database}
\label{fig:dbschema}
\end{figure}

\end{example}

In this setting, IT developers might consider the presence of many surrogate keys in the database schema as a sufficient guarantee of its quality and may be tempted to not encode natural keys, as they are perceived as complex to manage at the application level, which opens the door to redundancy. 
Consequently, the unicity enforced by surrogate keys might be artificial, referred to as \textit{artificial unicity} in this paper. This is rather counterintuitive, since normalization associated to keys and foreign keys is supposed to prevent redundancy.

\begin{example}
Assume that the owner of \textit{Perfect Pet} hired a data-science consultant to leverage its existing data and better understand their patients' profiles. Table \ref{tab:db} includes an extract from their database. 
When looking at the relation \texttt{Animal}, she observes that the values associated to attributes $\langle$ \texttt{species, breed, name, gender, dob} $ \rangle$ are duplicated on tuples 1, 4, and 7, but the values of the relation’s natural key: \texttt{id\_microchip} (611, 616, and 620) are distinct on these three tuples.
However, tuples associated to these id in relation \texttt{Microchip} share the same value on attribute \texttt{number}, the natural key of the relation (744982769), indicating that they describe the same microchip, thus that the three tuples from relation \texttt{Animal} indeed describe the same dog, despite their unique values on the key \texttt{id\_animal}. The unicity of this relation's data is qualified as "artificial".
\end{example}

\begin{table*}[bt]
\caption{Extract of the clinic's database}
	\begin{center}
	\centering 
	\ContinuedFloat
		\begin{subtable}{1\textwidth}
		\centering
			\begin{scriptsize}
			\centering 
			\begin{tabular}{c|c|c|c|c|c|c|c|c|c|c}
			 & \textbf{id\_animal} & \textbf{species} & \textbf{breed} & \textbf{name} & \textbf{id\_microchip} & \textbf{gender} & \textbf{dob}  & \textbf{weight} & \textbf{food} & \textbf{hash\_id}\\
			\hline
			1 & 744 & canine & greyhound & Gunter & 611 & M & 15-07-2014 & 27 & Greyh. Mix & 227f1df55c\\
			2 & 746 & canine & labrador & Coco & 613 & F & - & 41 & H Large Dog & c59a2c0744\\
			3 & 747 & feline & siamese & Izzy & 614 & F & 18-01-2018 & 4.2 & Indoor Cat & 9d5faf7fa6\\
			4 & 749 & canine & greyhound & Gunter & 616 & M & 15-07-2014 & 25.8 & Urinary & ded5f4207a\\
			5 & 750 & feline & siamese & Izzy & 617 & F & 01-01-2018 & 4.1 & Indoor Cat & d33673d97d\\
			6 & 752 & canine & labrador & Coco & 619 & F & - & 41 & H Large Dog & 4745dd610e\\
			7 & 753 & canine & greyhound & Gunter & 620 & M & 15-07-2014 & 26 & Urinary & f41e45bf6e\\
			\hline
			\end{tabular}
			\subcaption{Animal}
			\end{scriptsize}
		\end{subtable}

		\begin{subtable}[c]{1\textwidth}
		\centering
			\begin{center}
			\begin{scriptsize}
			\begin{tabular}{c|c|c|c|c|c}
			 & \textbf{id\_appointment} & \textbf{id\_animal} & \textbf{date} & \textbf{time} & \textbf{main\_reason} \\
			\hline
			1 & 370 & 744 & 08-02-2022 & 10:00 & sick pet \\
			2 & 372 & 746 & 20-04-2022 & 15:00 & surgery\\
			3 & 373 & 747 & 10-05-2022 & 11:30 & injured pet\\
			4 & 375 & 749 & 20-07-2022 & 16:00 & annual visit\\
			5 & 376 & 750 & 20-07-2022 & 14:30 & annual visit\\
			6 & 378 & 752 & 16-08-2022 & 12:00 &surgery follow-up\\
			7 & 379 & 753 & 09-09-2022 & 17:00 & sick pet\\
			\hline
			\end{tabular}
			\subcaption{\texttt{Appointment}}
			\end{scriptsize}
			\end{center}	
		\end{subtable}
		
		\begin{subtable}[c]{1\textwidth}
		\centering
			\begin{center}
			\begin{scriptsize}
			\begin{tabular}{c|c|c|c}
			 & \textbf{id\_microchip} & \textbf{number} & \textbf{implant\_date} \\
			\hline
			1 & 611 & 744982769 & 09-04-2015 \\
			2 & 613 & 713356552 & 18-11-2018 \\
			3 & 614 & 576918144 & 23-06-2018 \\
			4 & 616 & 744982769 & 04-09-2015 \\
			5 & 617 & 576918144 & 23-06-2018 \\
			6 & 619 & 713356552 & 18-11-2018 \\
			7 & 620 & 744982769 & 09-04-2015 \\
			\hline
			\end{tabular}
			\subcaption{\texttt{Microchip}}
			\end{scriptsize}
			\end{center}	
		\end{subtable}
		
	\end{center}
\label{tab:db}
\end{table*}

Artificial unicity and associated underlying data quality issues could accumulate over the years and may go unnoticed as long as it does not impact the delivery of the digital applications nor the operations of the SME.
Moreover data quality is known to impact machine learning (ML) models' performance  \cite{lee2021survey, budach2022effects, zha2023data}, 
especially redundancy by altering the accuracy of simple statistics such as median, average, distribution, or frequency.

\begin{example}
\label{ex:sql1}
Assume that the data scientist writes the following SQL query to get a dataset in order to learn a model predicting the number of appointments for animals:

\begin{small}
\begin{verbatim}
SELECT id_microchip, species, breed, gender, COUNT(*) as nb_apt
FROM Animal an
JOIN Appointment ap ON an.id_animal = ap.id_animal
GROUP BY ap.id_animal
\end{verbatim}
\end{small}
She will get 7 tuples 
due to the artificial unicity in the database, instead of 3. 
Instead of being represented once and associated to three appointments, for instance, the dog Gunter from example 1.2 appears three times and its number of appointments reduced to one each time.
Any model learned from this data would be biased. 
\end{example}

In the context of analytics, dirty data can be cleaned either at the source (while still in the relational format), or after it has been extracted into a dataset for a specific task.
The second option is the most popular as it appears to be less time-consuming.
However, the generalized presence of artificial unicity prevents the detection and resolution of duplicates. 
We argue that artificial unicity has to be removed at the source before any regular cleaning technique can be applied. 
Any attempt to remove artificial unicity on the answer set of a SQL query is likely to fail.
The main reason comes from the spread of artificial unicity induced by surrogate keys to the whole database through the keys-foreign keys join path. \\

In this setting, we now state the main problem we address in this paper.

\emph{Problem statement:} Given an operational relational database devised mainly with surrogate keys, how do we assess whether the database contains artificial unicity?
If it does, how can we extract a reduced, redundancy-free, and normalized database?\\

Solving this problem draws from different disciplines that have been studied for years, from integrity-constraint elicitation, also known as data profiling \cite{DBLP:journals/vldb/AbedjanGN15} including functional dependencies, inclusion dependencies or keys \cite{DBLP:journals/jiis/MarchiLP09, DBLP:journals/jiis/JiangN20,DBLP:journals/tkde/CaruccioDNP21, chu2013discovering}, to entity resolution \cite{DBLP:journals/jdiq/DraisbachCN20, christen2012data, barlaug2021neural, bhattacharya2007collective, gokhale2014corleone, wang2012crowder, wu2020zeroer}, and database reverse engineering or re-engineering \cite{DBLP:journals/ijcis/PetitTK95}. 
Despite their sophistication and robustness, these techniques still encounter obstacles to their adoption by SMEs such as their high computational cost, difficulty to scale up to large amounts of data or to scale to real-life database characteristics. 
From our experience, we believe that an interactive holistic framework is required to address the problems posed by artificial unicity.\\

\emph{Paper contribution:}
This paper, grounded in real-world applications, defines the concept of artificial unicity and proposes \textit{RED2Hunt (RElational Databases REDundancy Hunting)}, a human-in-the-loop framework for identifying hidden redundancy and, if problems occur, cleaning relational databases implemented with surrogate keys. 
We first define the central and intricate notion of artificial unicity and then the RED2Hunt framework to address it. 
We rely on simple abstractions easy to visualize based on the so-called \emph{redun\-dancy profile} associated to some relations and the notion of \emph{attribute stability}. Quite interestingly, those profiles can be computed very efficiently in quasi-linear time. We have devised different metrics to guide the domain expert and an actionable framework to generate new redundancy-free databases. The proposed framework was implemented on top of PostgreSQL. 
From the publicly available IMDB database, we have generated synthetic databases, implementing different redundancy scenarios, on which we tested RED2Hunt to study its scalability.
RED2Hunt has also been tested on operational databases implemented with surrogate keys. Lessons learned from these real-life applications are discussed.\\

\emph{Paper organization:}
First, we remind the reader of some database concepts and notations in section 2 and formally define the concept of artificial unicity in section 3. Then, we introduce the \textit{RED2Hunt} framework in section 4. 
Section 5 is dedicated to the implementation and application of the framework.
Section 6 presents a review of related work, and section 7 concludes this paper.


\section{Preliminaries}

In this section, we introduce the notations to be used throughout the paper. It is assumed that the reader is familiar with database notations \cite{levene2012guided}.\\

Let $U$ be a set of attributes and $D$ a set of constants such that $U \cap D = \emptyset$.  
A relation schema $R$ is defined over $U$, with $R \subseteq U$. 
A tuple $t$ over $R$ is a function from $R$ to $D$. 
A relation $r$ over $R$ is a finite set of tuples over $R$. 
A database schema $\mathcal{R}$ over $U$ is a finite set of relation schemas over $U$ and a database $d$ over $\mathcal{R}$ is a set of relations $r$ over $R$, for every $R \in \mathcal{R}$. 
Let $X \subseteq U$  and $t $ a tuple over $U$. $t[X]$ is the restriction of $t$ to $X$.
Let $r$ be a relation over $R$ and $X \subseteq R$. 
The projection of $r$ over $X$, denoted $\pi_{X}(r)$, is defined as usually $\pi_{X}(r) = \{t[X] | t \in r\}$. We adopt set semantics for relational-algebra expressions.
The size of a finite set $E$ is denoted by $\card{E}$.

A similarity $\simeq$ is a binary relation over $D$, denoted $v_1 \simeq v_2$ for $v_1, v_2 \in D$. 
A similarity is supposed to be reflexive, symmetric, and transitive.

\paragraph{Functional dependencies}

Let $R$ be a relation schema, $X \subseteq R$ and $A \in R$.
A functional dependency over $R$ is denoted by $R: X \rightarrow A$. 
Let $r$ be a relation over $R$. $R: X \rightarrow A$ is satisfied in $r$ ($r \vDash X \rightarrow A$) if and only if:
$\forall t_{1}, t_{2} \in r$ if $\forall B \in X: t_{1}[B] = t_{2}[B]$ then $t_{1}[A] = t_{2}[A]$. 

A classical measure to approximate the satisfaction of FDs is known as the $g_3$ measure \cite{kivinen1995approximate}.
It corresponds to the smallest proportion of tuples to be removed form the relation $r$ for $X \rightarrow A$ to hold in $r$. 
More formally,we have: 
$$ g_3(r, X \rightarrow A) = 1 - \frac{\mathtt{max}(\card{r'} \mid r' \subseteq r, r' \models X \rightarrow A)}{\card{r}} $$

For the sake of clearness, we only consider equality even if functional dependencies extends to diverse similarity measures. 

\paragraph{Closure, key and foreign key}

Let $F$ be a set of functional dependencies over $U$ and $X \rightarrow Y$ a functional dependencies over $U$.
 $X \rightarrow Y$ is implied by $F$, known as the implication problem, is denoted by $F \models X \rightarrow Y$.

Let $X \subseteq U$. The closure of $X$ with respect to $F$, denoted $X_F^+$, is defined by  $X_F^+ = \{A \in U | F \models X \rightarrow A\}$. 
$X$ is a superkey of $F$ in $R$ if $X^{+}_{F} = R$ and just a key if no proper subset of $X$ is still a key. 
We may have several candidate keys in a given relation schema, among which one is chosen to be the \emph{primary key (PK)}. A PK can be a \emph{natural key (NK)} or a \emph{surrogate key (SK)}. Surrogate keys always pertain to only one attribute.

A foreign key is the left hand side of an inclusion dependency whose right hand side is a key (cf. example \ref{fig:dbschema}). 
We denote by $\mathcal{I}$ the set of associated inclusion dependencies, and by $\mathcal{I}_{unary}$ the set of induced unary inclusion dependencies. 
A foreign key is said to be surrogate whenever it points to a surrogate key. 

\paragraph{Graphs}

Every relational database schema can be represented as an oriented graph $G = (V, E)$ where $V$ is the set of relations and $E$ is the set of edges induced by foreign keys. 
Let $u, v \in V$, $(u,v) \in E$ if there exists a foreign key from some attribute(s) of $u$ pointing to some attribute(s) of $v$. 

$d^{-}(s)$ and $d^{+}(s)$ are the indegree and outdegree of node $s$. In relational terms, $d^{-}(s)$ is the number of foreign keys leaving from $s$ and $d^{+}(s)$ is the number of foreign key entering into $s$.

\section{ARTIFICIAL UNICITY DEFINITION}

Artificial unicity qualifies the redundancy present in relational databases devised with surrogate keys and their associated surrogate foreign keys. 
Artificial unicity turns out to be tricky to define and somehow counterintuitive as a surrogate key cannot contain duplicate values but at the same time, several values of a surrogate key may refer to the very same information. 
To define the concept of artificial unicity, we focus on specific relations which have the following characteristics:
\begin{itemize}
\item They have at least one surrogate key. We assume for the sake of simplicity that none of them have more than one surrogate key.
\item They do not have any  natural key specified in the DDL of the DBMS, even if a so-called \emph{undeclared natural key} may exist. 
\end{itemize}
This leads us to define the notion of \textit{potential key}, applicable to all relations having a surrogate key:

\begin{definition}
Let $r$ be a relation over $R$, $SK \in R$ a surrogate key, and $X \subseteq R \setminus \{SK\}$.\\
$X$ is a \textbf{potential key} of $R$ if:
\begin{itemize}
\item $X = NK$ where $NK$ is an undeclared natural key of $R$,
\item $X = R \setminus \{SK\}$ otherwise.
\end{itemize}
\end{definition}

For those relations, a potentiel key may take two forms: either as a natural key (if it exists), or as all the attributes but the surrogate key. For the latter, the intuition is to say that it does not make much sense to have duplicated values for all the attributes of the relation, except the surrogate key. This suggests that those attributes form a candidate key.

It is worth noting that we do not make any assumption about the data quality of a potential key in a relation, for example duplicated values or dirty data may occur. 

The elicitation of potential keys will be described in section 4. 

\begin{example}
In our running example, 
the attribute \texttt{number} in \texttt{Microchip}, the combination of the three attributes \texttt{date}, \texttt{time} and \texttt{id\_animal} in \texttt{Appointment}, and \texttt{id\_microchip} in \texttt{Animal}, are potential keys. 
\end{example}

We now provide an inductive definition of a central notion to define artificial unicity, called \emph{artificially unique duplicates} within a relation with respect to the whole database, as follows:

\begin{definition}
Let $d$ be a database over $\mathcal{R}$, $r \in d$ a relation over $R \in \mathcal{R}$, $\mathcal{I}_{unary}$ the set of unary inclusion dependencies, 
$SK \in R$ a surrogate key of $R$, $X$ a potential key of $R$, and $t_1, t_2 \in r$.\\
$t_1$ and $t_2$ are \textbf{artificially unique duplicates in $d$ with regard to $SK$}, denoted $t_1[SK] \equiv^r_{SK} t_2[SK]$ (or simply $t_1[SK] \equiv t_2[SK]$ when clear from context), if for every $A \in X$:

\begin{itemize}
\item either $\nexists R[A] \subseteq S[K] \in \mathcal{I}_{unary}$ and $t_1[A] \simeq t_2[A]$,
\item or $\exists R[A] \subseteq S[K] \in \mathcal{I}_{unary}, s \in d$ over $S$, and either $t_1[A] = t_2[A]$ or $t_1[A] \equiv^s_{K} t_2[A]$.
\end{itemize}
\end{definition}

\begin{figure}[h]
	\caption{Artificial unicity propagation}
	\label{fig:au_propag}
	\centering
	\includegraphics[width=8cm]{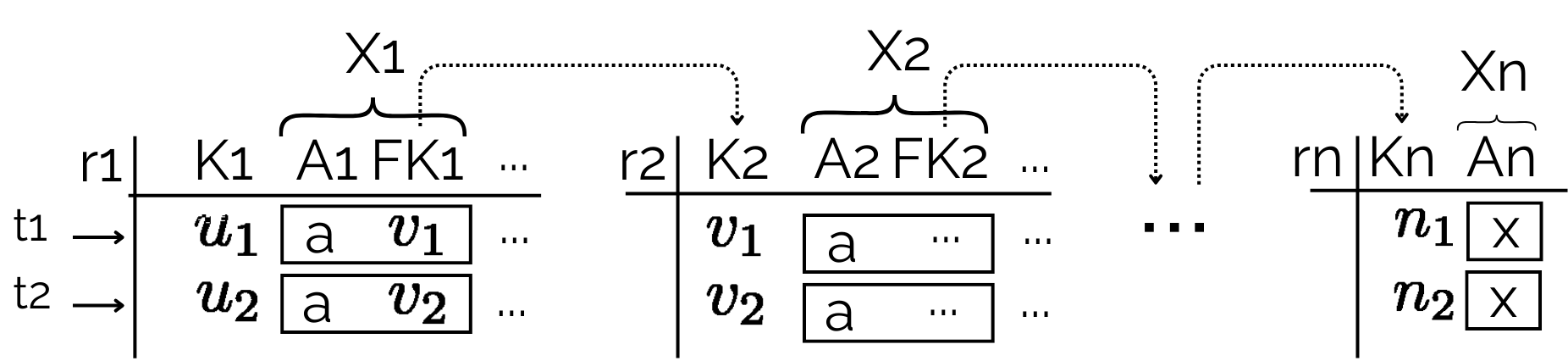}
\end{figure}

This definition allows to propagate artificial unicity through join paths as illustrated in figure \ref{fig:au_propag}.

\begin{example}
In our running example, consider relation \texttt{Animal} , $t_1[id\_animal] \equiv t_4[id\_animal] \equiv t_{7}[id\_animal]$ (i.e. $744 \equiv 749 \equiv 753$), since the potential key of animal \texttt{id\_microchip} is a foreign key with equivalent values in \texttt{Microchip}: $611 \equiv 616 \equiv 620$.
\end{example}

Finally, we can define \emph{artificial unicity} at the level of a relation, as follows:

\begin{definition}
Let $d$ be a database over $\mathcal{R}$, $r \in d$ a relation over $R \in \mathcal{R}$, and $SK$ its surrogate key.\\
$r$ suffers from \textbf{artificial unicity induced by surrogate keys} in $d$ if there exists at least two tuples $t_1, t_2 \in r$ such that $t_1$ and $t_2$ are artificially unique duplicates in $d$ with respect to $SK$.
\end{definition}

\begin{example}
Following our previous example, \texttt{Microchip} and \texttt{Animal} suffer from artificial unicity induced by surrogate keys.
\end{example}

In the next section we introduce RED2Hunt, a human-in-the-loop framework devised to remove artificial unicity and clean associated data quality problems from an operational relational database. 

\section{RED2HUNT FRAMEWORK}

\begin{figure*}[h]
	\caption{Overview of RED2Hunt}
	\label{fig:process_overview}
	\includegraphics[width=16cm]{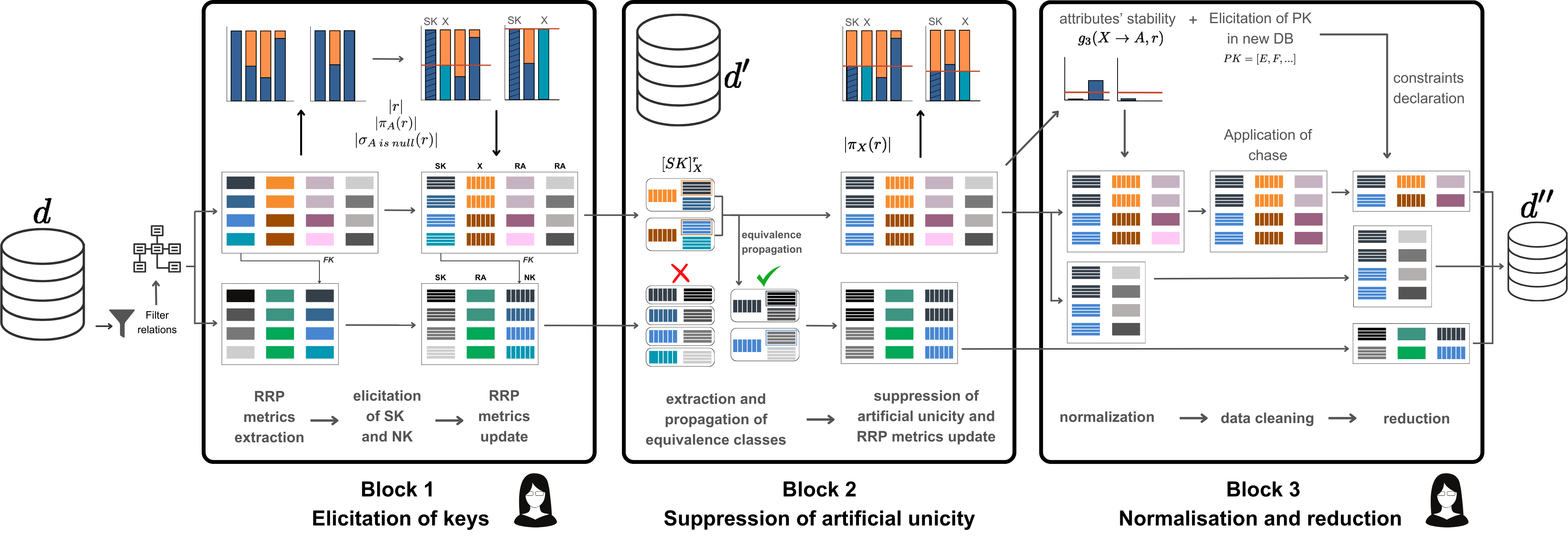}
\end{figure*}

RED2Hunt's purposes are, given an operational database : 1) to quantify its artificial unicity and 2) create a redundancy-free version to be used for analytical or ML needs. 
It is composed of three main blocks as represented in Figure \ref{fig:process_overview}:
\begin{enumerate}
\item Elicitation of keys: Meta-data characterizing the relations are extracted and displayed as visuals to guide the domain expert through the elicitation of all kinds of keys.
\item Suppression of artificial unicity: A new database is built in which the values of surrogate keys inducing artificial unicity are corrected, allowing a global assessment of the artificial unicity and its suppression.
\item Normalisation and reduction: The new database is cleaned of remaining data quality issues by the decomposition of some relation schemas and duplicates resolution.
\end{enumerate}

Each block described in Figure \ref{fig:process_overview} is illustrated with \textit{Perfect Pet}'s running example.
Since we are dealing with operational databases, we first enable the filtering of tables that are of most interest to the expert.  
Block 1 and 3 require interaction with domain experts. 

\subsection{Elicitation of keys}

Declared constraints on tables such as primary key, foreign key, unique and not null constraints are easily retrieved from the data definition language (DDL) of any DBMS.
For each selected relation, such keys have to be identified whenever they exist. 
For undeclared keys, instead of relying on data profiling techniques based on keys or functional-dependency discovery, we propose to the domain expert a new-yet-simple notion, the so-called \emph{redundancy profiles} of a relation to guide her in this elicitation process.

\subsubsection{Relation redundancy profiles}

\begin{definition}
Let $r$ be a relation over $R$ and $A \in R$.\\
The \emph{attribute redundancy profile} of $A$ in $r$, denoted by $ARP(A,r)$, is defined by $ARP(A,r) = \langle dv, nv, ov \rangle$ with $dv$ the number of distinct values,  $nv$ the number of null values and  $ov$ the number of duplicate values of $A$ in $r$, computed as $ov = |r| - dv - nv$.
\end{definition}

Quite clearly, such information can be computed in quasi linear time, and its extraction is straightforward in SQL.
Every ARP can be visualized as a staked bar with a color associated to each value. 
$dv$ is represented by blue, $nv$ by green and $ov$ by orange (see Figure \ref{fig:pp_rrps}).
The previous definition extends easily to relation as follows.

\begin{definition}
The \emph{relation redundancy profile} of $r$, denoted by $RRP(r)$, is the collection of attribute redundancy profiles of its attributes, i.e. $RRP(r) = \bigcup_{A\in R} ARP(A,r)$. \end{definition}

To account for the constraints defined in the DDL, each attribute belongs to one of the three categories: declared Keys (K), Foreign-Keys (FK), and Remaining Attributes (RA).
A simple and intuitive visual representation can be associated to every RRP, providing a summary of its attributes, taken one by one: on the far left, we plot the attributes defined as keys, on the far right, the attributes that are foreign keys, and in the middle, all remaining attributes sorted according to their number of distinct values in descending order.
On top of the graph is plotted a horizontal blue line representing the size of the relation, making the RRP visualizations a great revealer for the domain expert to identify attributes that are keys or tend to be keys but are not declared as such in the DDL.

\begin{example}
Continuing our example, we depict in Figure \ref{Fig:pp_rrp_a1} the initial visualization of the redundancy profile of \texttt{Animal}, recall that \texttt{id\_animal} is declared as a primary key in \texttt{Animal}. The associated stacked bar is therefore a single blue bar with a 'K' annotation above it.
\texttt{hash\_id} and \texttt{id\_microchip} appears to be undeclared keys.
\end{example}

\subsubsection{Elicitation of surrogate keys}

The goal of this step is to identify surrogate keys. 
Since they always pertain to a single attribute, the RRP is an excellent abstraction for the expert. 
There are several identification scenarios:
\begin{itemize}
\item A surrogate key has been declared as such in the DDL [e.g. \textsf{uuid} attribute in PostgreSQL). 
\item A key (primary key or unique non-null) has been defined on a single attribute and its values have no semantic meaning for the domain expert.
\item An attribute appears as a key (or almost a key) but has not been declared as such in the DDL (positioned in the RRP right after the keys or is a foreign-key), and its values have no semantic meaning for the domain expert. 
\end{itemize}
These surrogate key attributes are tagged as $SK$ and then reflected on the RRPs. 

\begin{example}
Continuing our  example, the expert, from her background knowledge, could determine which of these \texttt{Animal}'s attributes are indeed surrogates keys. From a data sample, she identifies \texttt{hash\_id} as being a surrogate key, as well as the primary key (defined in the DDL).
\end{example}

\subsubsection{Elicitation of natural keys}

Assessment of artificial unicity relies on the identification of the potential keys, which correspond to either natural keys in relations wherever they exist or to the reamining set of attributes without any surrogate key in relations. 
Therefore, elicitation of natural keys within the selected relations is clearly one of the most crucial steps of our proposition. 

We favor an expert-based approach to avoid the known pitfalls of methods for discovering minimal keys in a relation \cite{milo1999efficient}, notably the high number of false positives and false negatives.
Instead, the domain expert is asked to identify a single attribute or a combination of attributes that may be a \emph{natural key} of each selected relation.

In the type of databases we target, it is assumed that some form of normalization has taken place, resulting in a set of SQL tables; each conveying a semantic meaning, often carried by its name and meaningful to the domain expert. As a consequence, it is likely that at least one natural key should exist.
RRPs combined with visualization of data samples again offer an excellent means to explore possible natural keys. 

Whenever the expert identifies a natural key in a relation, if it comprises more than one attribute, a new column is created in its redundancy profile. 
A horizontal red bar is also drawn at the number of unique values of this natural key. Everything above this red line is likely to be redundant. 
The visual difference between the two horizontal lines (blue and red) offers an initial estimate of the redundancy existing in this relation.

\begin{figure*}
\caption{Evolution of RRPs for \textit{Perfect Pet}'s database}
\label{fig:pp_rrps}

        \begin{subfigure}[t]{0.3\textwidth}
            \raggedleft
            \includegraphics[width=5cm, keepaspectratio]{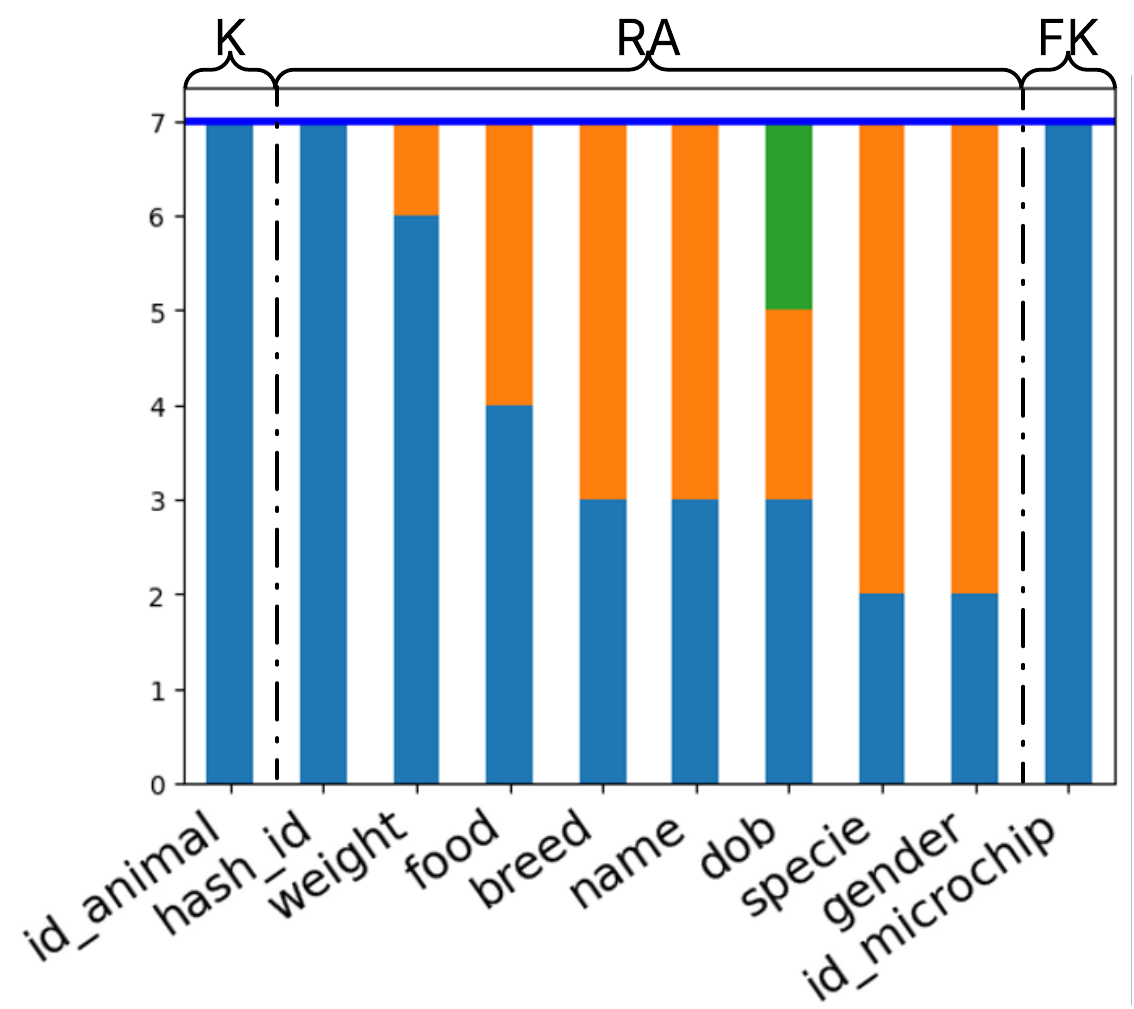}
            \caption{RRP(Animal) - initial}
            \label{Fig:pp_rrp_a1}
        \end{subfigure}
        \hspace{0.005\textwidth}
        \begin{subfigure}[t]{0.3\textwidth}
            \raggedleft
            \includegraphics[width=5cm, keepaspectratio]{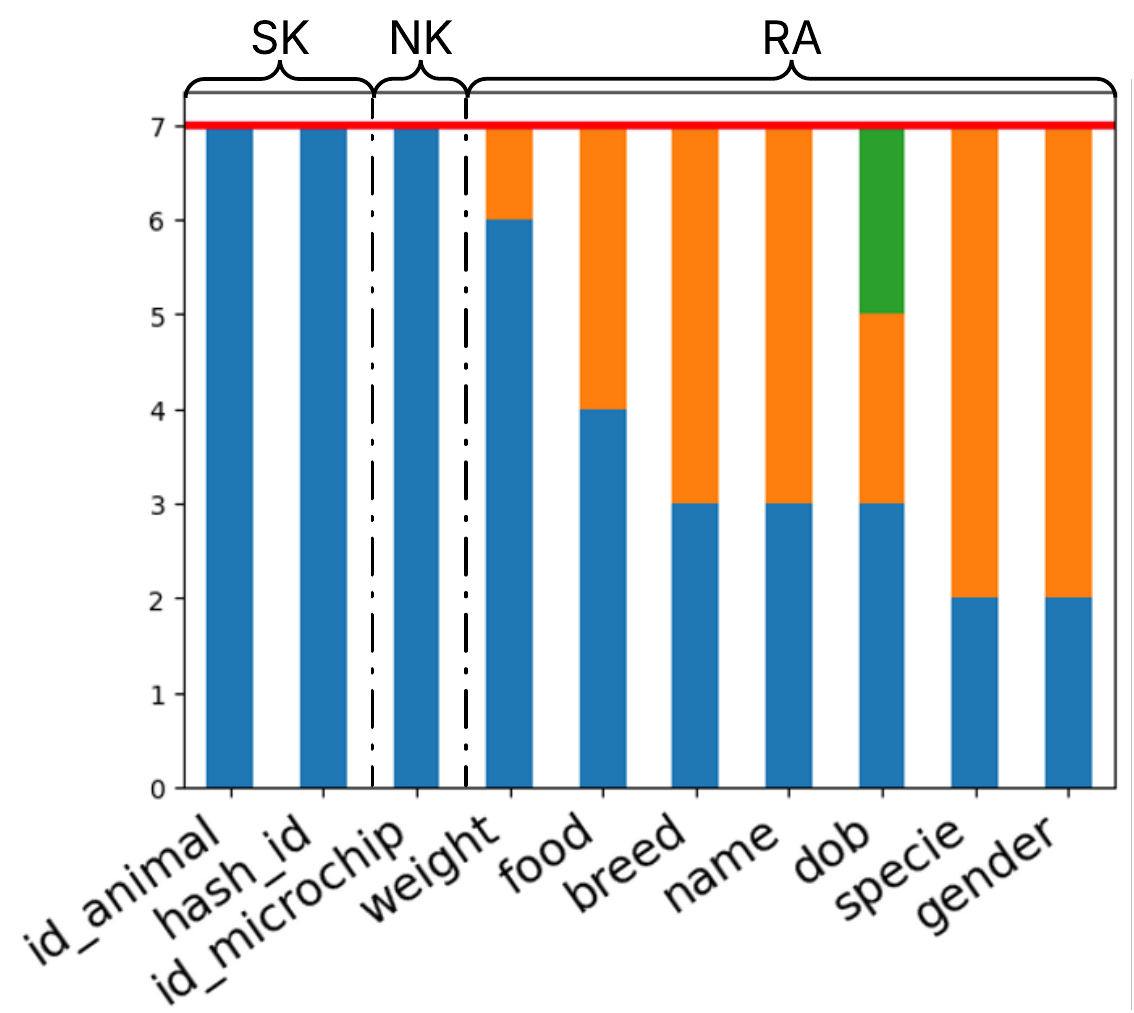}
            \caption{RRP(Animal) - end Block 1}
            \label{Fig:pp_rrp_a2}
        \end{subfigure}
        \hspace{0.005\textwidth}
        \begin{subfigure}[t]{0.3\textwidth}
           \raggedleft
            \includegraphics[width=5cm, keepaspectratio]{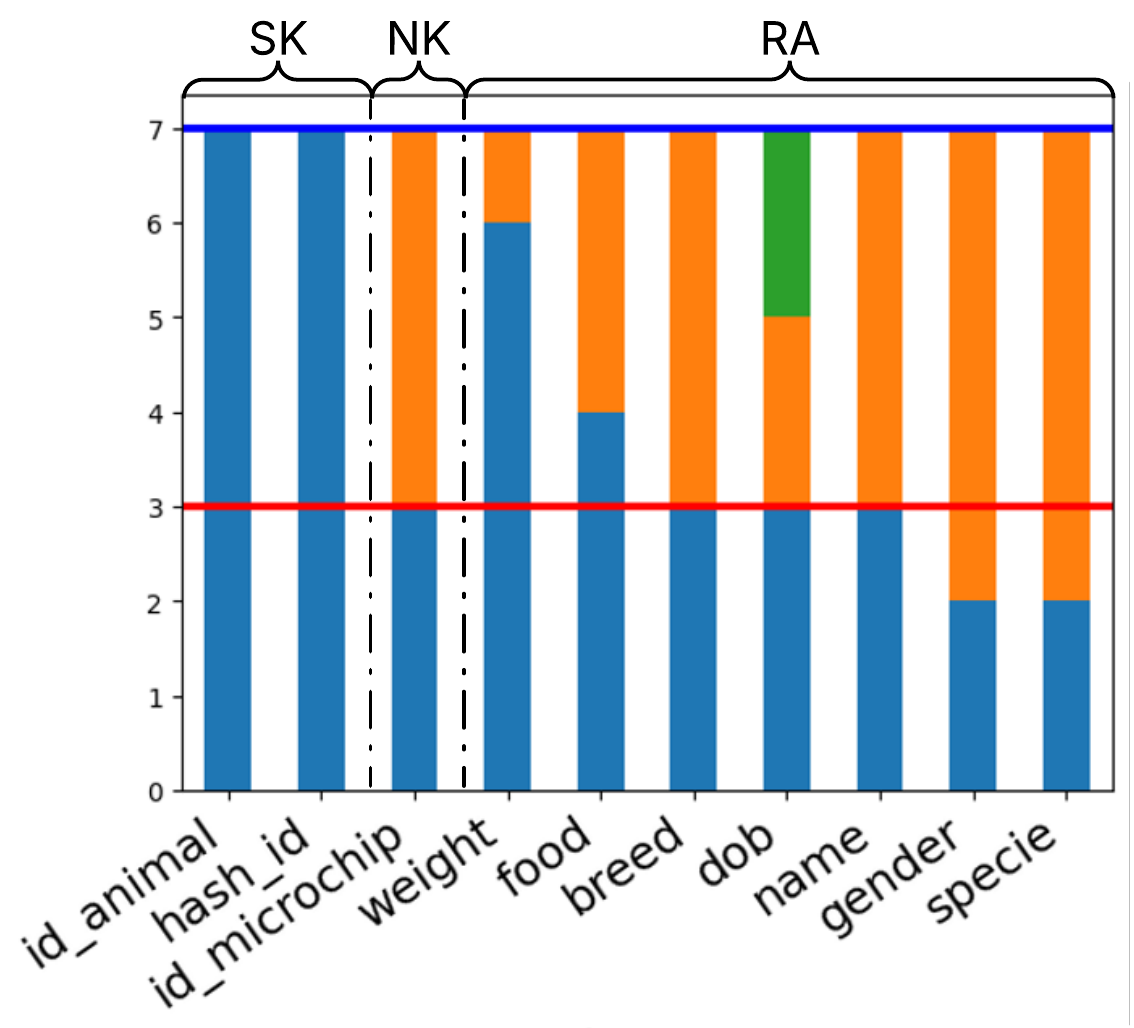}
            \caption{RRP(Animal) - end Block 2}
            \label{Fig:pp_rrp_a3}
        \end{subfigure}

        \begin{subfigure}[t]{0.3\textwidth}
            \raggedleft
            \includegraphics[width=4.5cm, keepaspectratio]{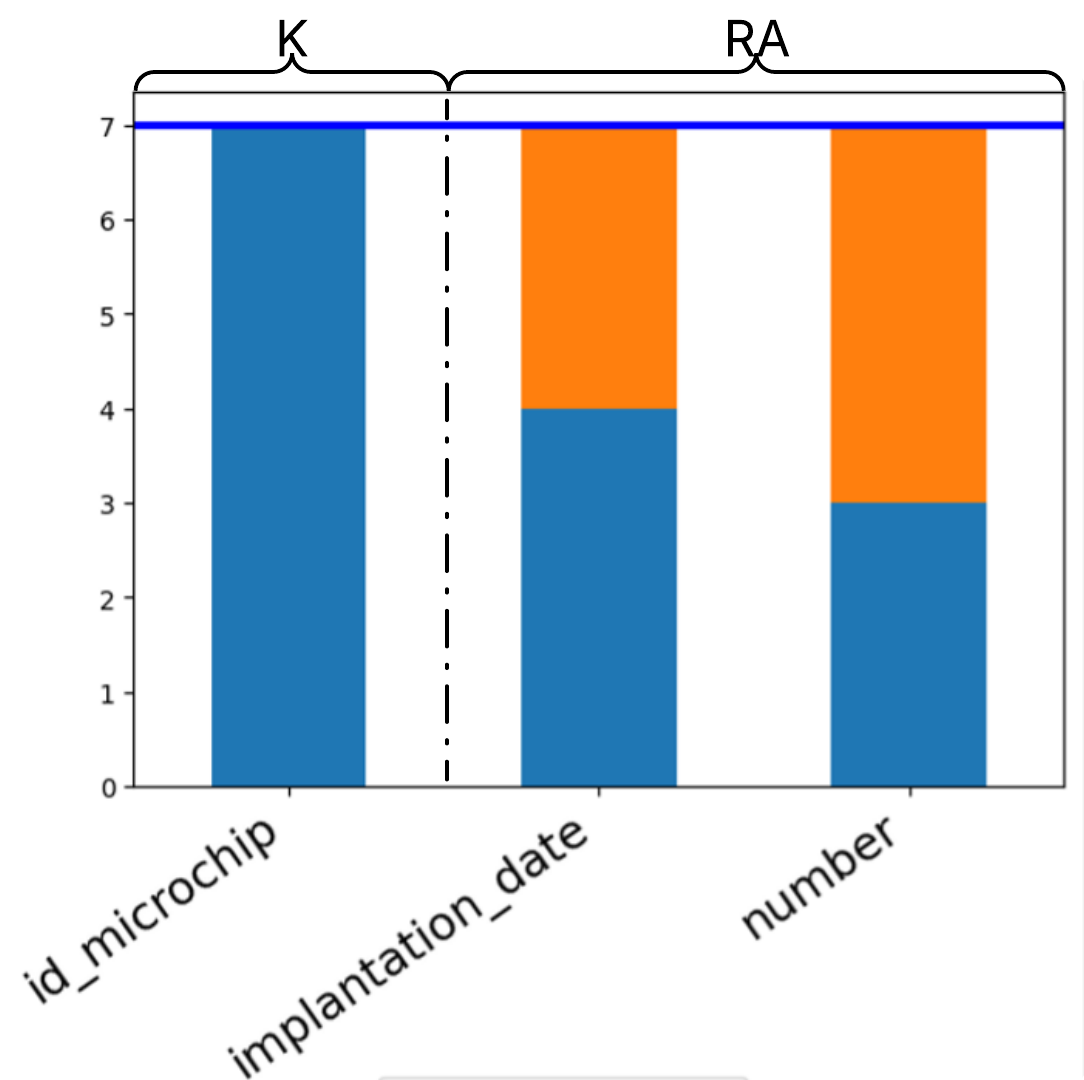}
            \caption{RRP(Microchip) - initial}
            \label{Fig:pp_rrp_m1}
        \end{subfigure}
       \hspace{0.005\textwidth}
        \begin{subfigure}[t]{0.3\textwidth}
            \raggedleft
            \includegraphics[width=4.5cm,  keepaspectratio]{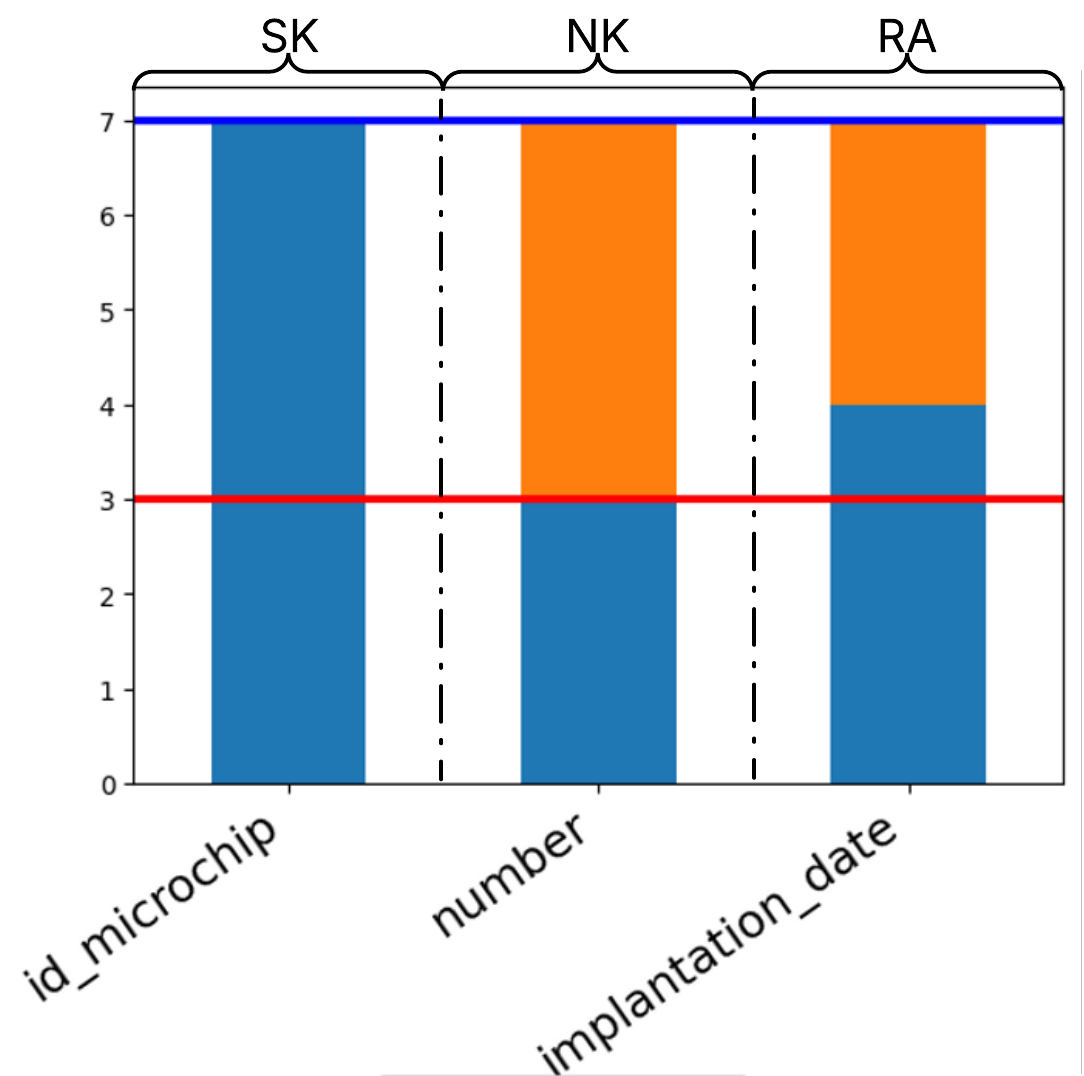}
            \caption{RRP(Microchip) - end Block 1}
            \label{Fig:pp_rrp_m2}
        \end{subfigure}
       \hspace{0.005\textwidth}
        \begin{subfigure}[t]{0.3\textwidth}
            \raggedleft
            \includegraphics[width=4.5cm, keepaspectratio]{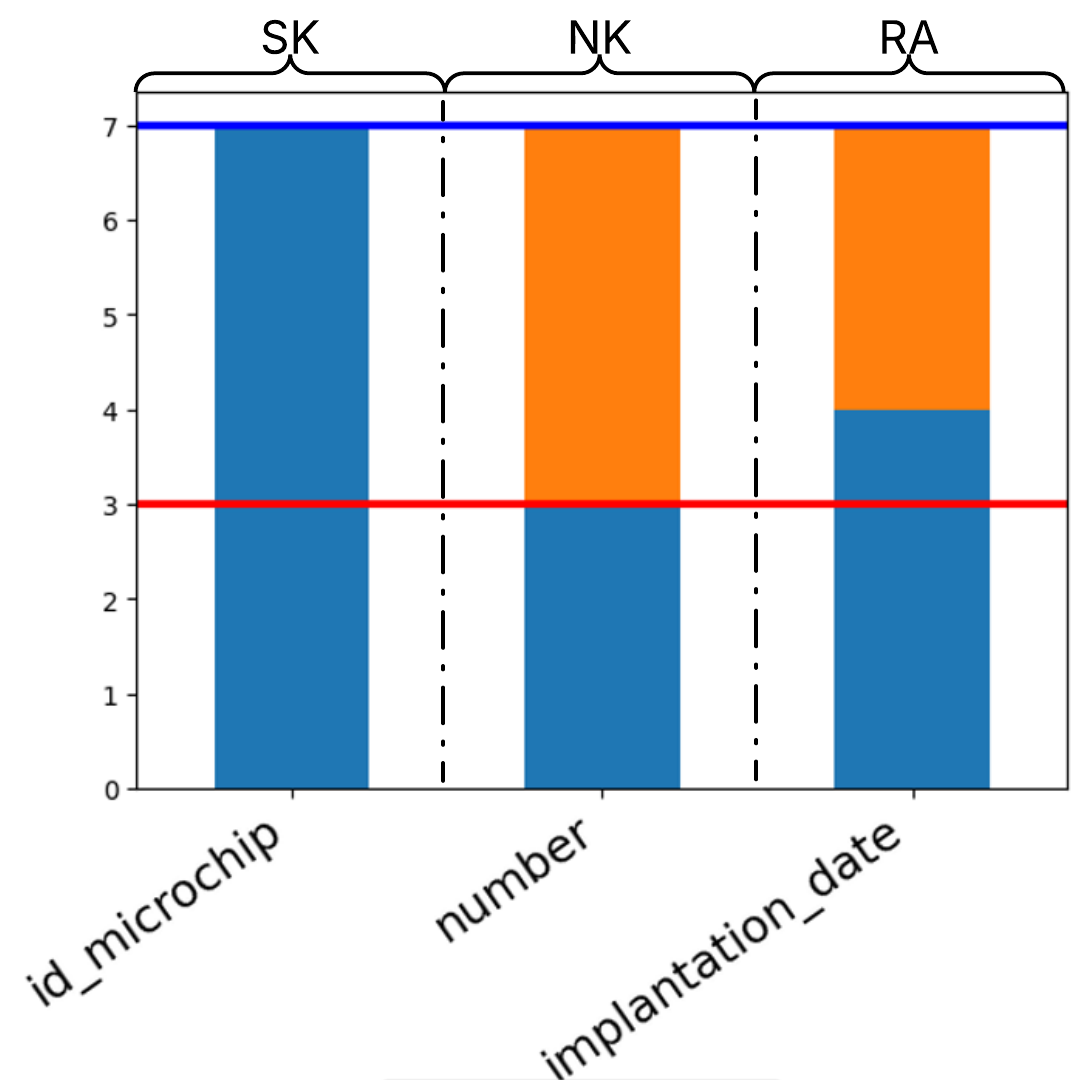}
            \caption{RRP(Microchip) - end Block 2}
            \label{Fig:pp_rrp_m3}
        \end{subfigure}

         \begin{subfigure}[t]{0.3\textwidth}
            \raggedleft
            \includegraphics[width=5cm, keepaspectratio]{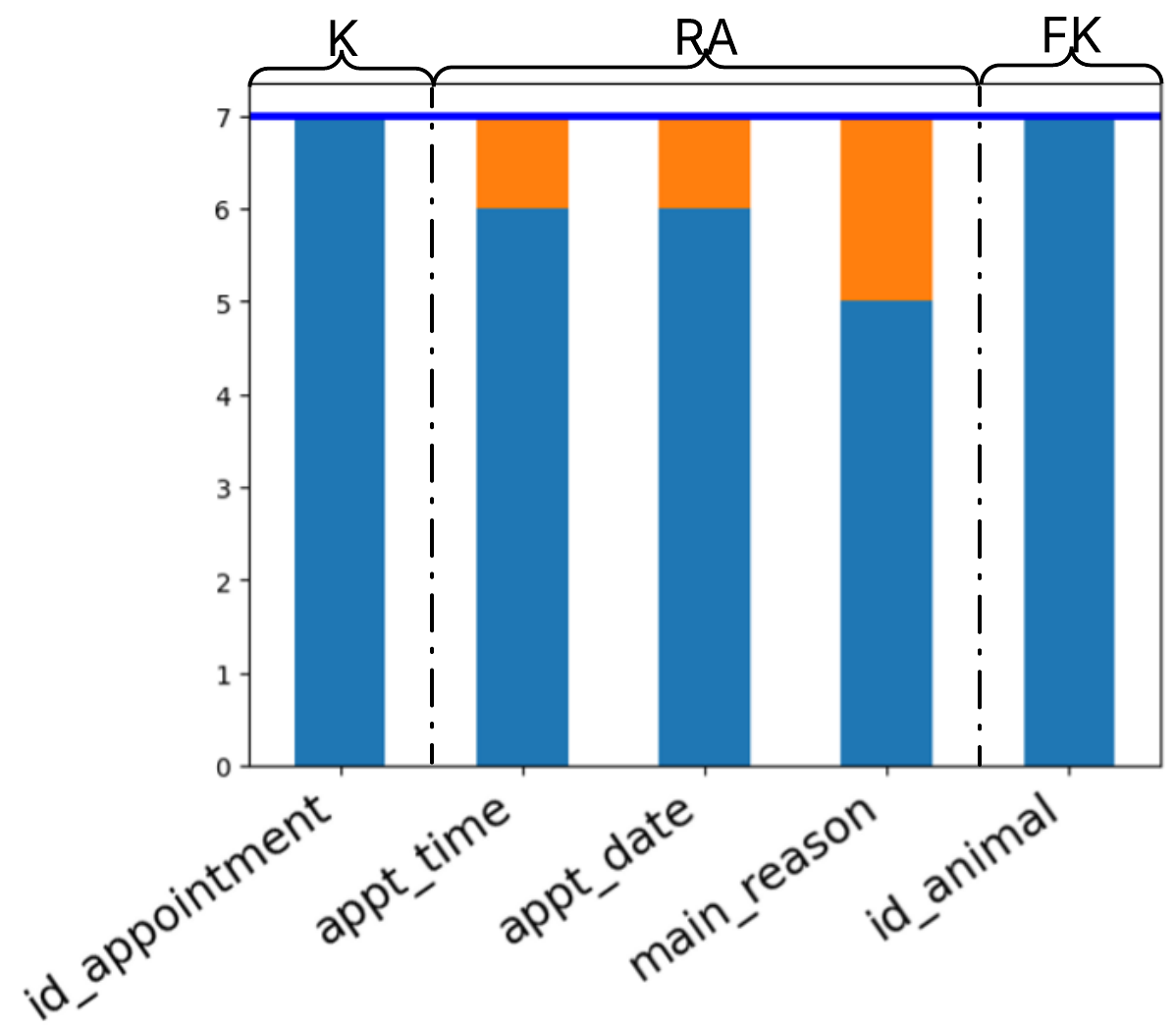}
            \caption{RRP(Appointment) - initial}
            \label{Fig:pp_rrp_ap1}
        \end{subfigure}
        \hspace{0.01\textwidth}
        \begin{subfigure}[t]{0.3\textwidth}
            \raggedleft
            \includegraphics[width=5cm, keepaspectratio]{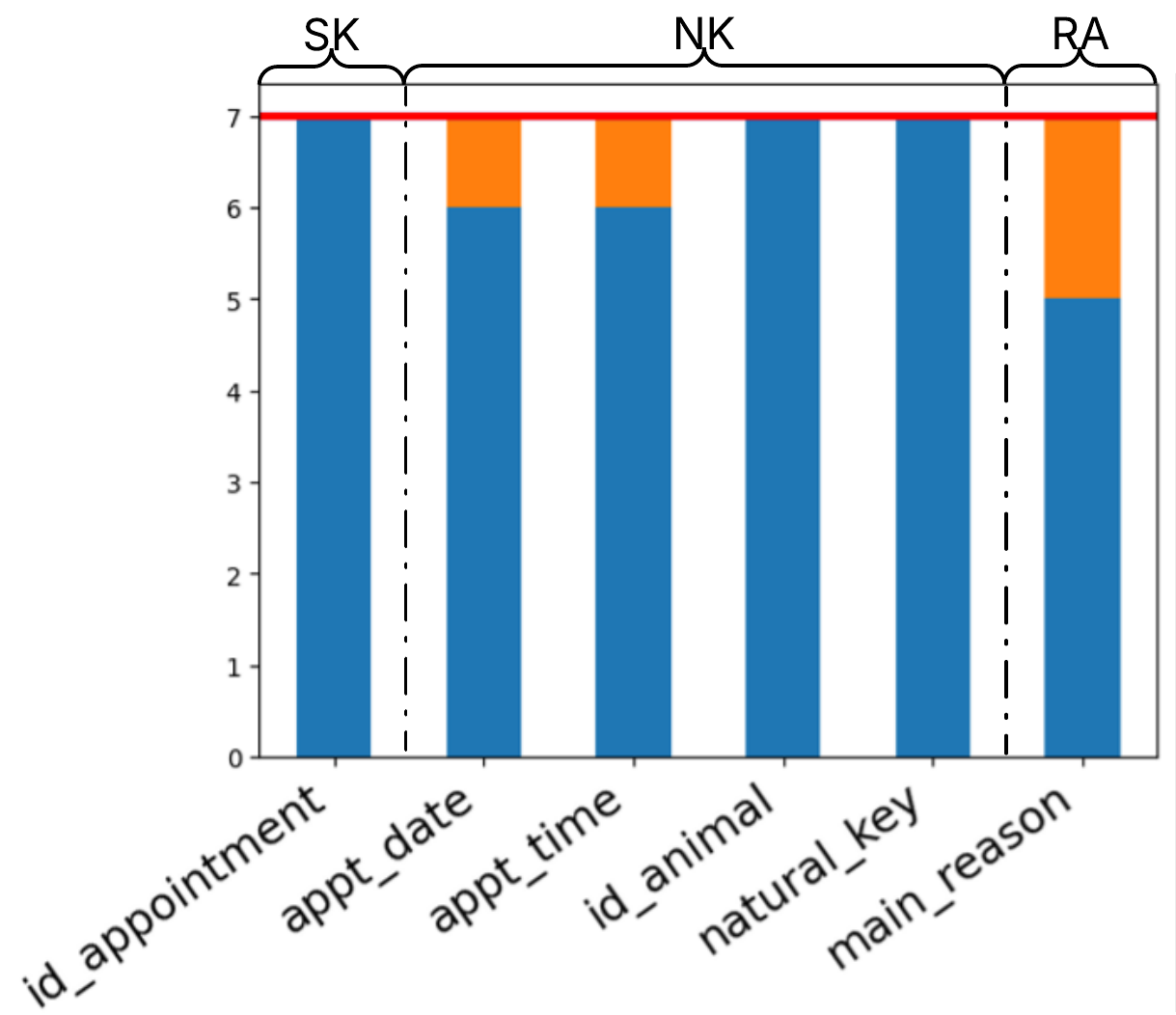}
            \caption{RRP(Appointment) - end Block 1}
            \label{Fig:pp_rrp_ap2}
        \end{subfigure}
        \hspace{0.01\textwidth}
        \begin{subfigure}[t]{0.3\textwidth}
            \raggedleft
            \includegraphics[width=5cm, keepaspectratio]{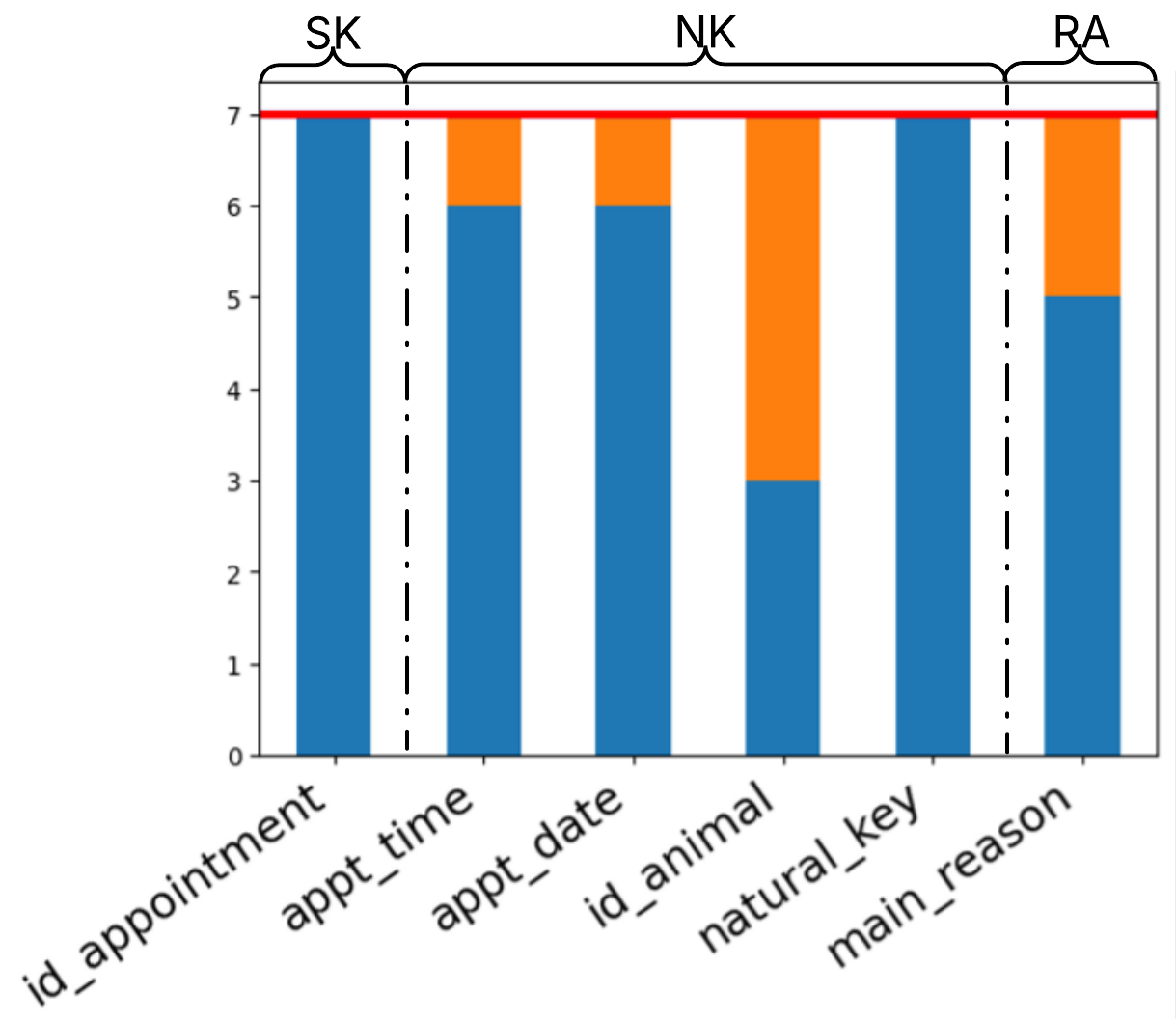}
            \caption{RRP(Appointment) - end Block 2}
            \label{Fig:pp_rrp_ap3}
        \end{subfigure}
        
                \vspace{0.4cm}
        \begin{subfigure}[t]{0.8\textwidth}
         	\includegraphics[width=8cm, keepaspectratio]{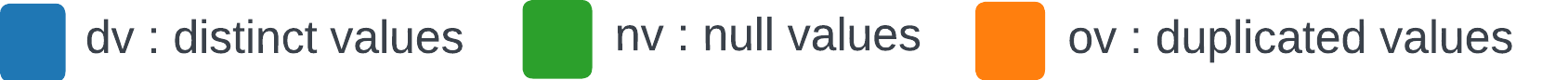}
        \end{subfigure}
        
\end{figure*}

\begin{example}
Assume that the domain expert at \textit{Perfect Pet} has identified the following natural keys: \texttt{id\_microchip} in \texttt{Animal}, \texttt{number} in \texttt{Microchip}, and \texttt{id\_animal + date + time} in \\ \texttt{Appointment}. 
Figures \ref{Fig:pp_rrp_a2}, \ref{Fig:pp_rrp_m2}, and \ref{Fig:pp_rrp_ap2} depict the visualization of the updated $RRP(\mathtt{Animal})$, $RRP(\mathtt{Microchip})$, and $RRP(\mathtt{Appointment})$ after elicitation of surrogate and natural keys. 
Quite easily, she observes a duplication problem in \texttt{Microchip} (Figure \ref{Fig:pp_rrp_m2}), while the two others look coherent so far.
\end{example}

The elicitation process can be applied several times with the domain expert until no more natural or surrogate keys appear.

\subsection{Artificial unicity suppression}

\subsubsection{Equivalence classes extraction}

If two tuples are artificial duplicates with respect to a surrogate key, then we say that the values on their surrogate keys are equivalent, and they belong to the same \textbf{equivalence class} with respect to the potential key.

\begin{definition}
Let $r$ be a relation over $R$, $t \in r$, $X \subset R$ its potential key, and $SK \in R \setminus X$ a surrogate key of $R$. 
The equivalent values of $t[SK]$ in $r$ with respect to $X$, 
denoted $[t[SK]]^r_{X}$, is defined by: 
$$[t[SK]]^r_{X} = \{
		t'[SK] \: | \: t' \in r \text{ and } t[SK] \equiv^r_{SK} t'[SK]\}$$
\end{definition}

This notion extends easily to all values of the surrogate key $SK$ in a relation.

\begin{definition}
The \textbf{equivalence classes of $SK$} in $r$ with respect to $X$, denoted $[SK]_X^r$, is defined by:
$$[SK]_{X}^r = \{ [t[SK]]^r_{X} \: | \: t \in r\}$$
\end{definition}

\begin{example}
In \textit{Perfect Pet}'s database, \texttt{Microchip} has a surrogate key \texttt{id\_microchip} and a potential key \texttt{number}. \
\begin{small}
$[id\_microchip]_{number}^{Microchip} = \{\{611, 616, 620\}, \{613, 619\}, \{614, 617\}\}$.
\end{small}
\end{example}

We introduce a map to clean a relation from its artificial unicity by cleaning its surrogate keys' values with respect to its potential key. 
For the sake of clarity, we keep only the minimum value on each equivalence class. 

\begin{definition}
Let $r$ be a relation over $R$, $SK \in R$ a surrogate key and $X \subset R \setminus \{SK\}$ its potential key. 
The map $clean\_SK(r, SK, X)$ returns a new relation over the same $R$ defined as follows: 
$clean\_SK(r, SK,X) = \{
	t \: | \: \exists t' \in r, t[R \setminus \{SK\}] = t'[R \setminus \{SK\}], t[SK] = min([t'[SK]]_X^r)
\}$  \\
\end{definition}

\subsubsection{Equivalence propagation}

The artificial unicity contained in the potential key has to be identified and removed before addressing the artificial unicity at the level of the relation, as illustrated in example 3.4. 
This suggests exploring every relation in a particular order, induced by the surrogate keys - surrogate foreign keys path. 

\begin{definition}
Let $s, r$ be two relations over $S, R$ resp., $FK \in S$ a surrogate foreign-key of $s$ pointing to a key $SK \in R$ of $r$ and $X \subset R$ the potential key of $r$.\\
$clean\_FK(s, FK, r, SK)$ returns a new relation $s'$ over $S$ as follows: \\
$clean\_FK(s, FK, r, SK) = \{
	t | \exists t' \in s, t[S \setminus \{FK\}] = t'[S \setminus \{FK\}], t[FK] = min([t'[FK]]_X^r)
\}$. 
\end{definition}

Since artificial unicity is spread throughout the database thanks to surrogate foreign keys, clearly no more propagation may occur when a relation with a declared natural key is reached. 
Such relations can be safely removed when chasing artificial unicity. 
Hence, since the propagation of artificial unicity within a database occurs only along the surrogate key-surrogate foreign key paths, we have to narrow the search and focus on a subset of relations and paths between them. 
From the oriented graph $(V, E)$ associated to a database, we focus on the subgraph induced by surrogate foreign keys only. 
We call this subgraph the \textbf{propagation graph}, denoted by $(V_{SK}, E_{SFK})$.

While classical graph associated to key-foreign key relationships may have cycles, this is likely to be less frequent with propagation graphs since many edges are not considered, i.e. those not associated to a surrogate foreign key, inducing a topological order among relations. 
In this paper we assume that the propagation graph is acyclic, leading to a linear ordering of the relations used to remove artificial unicity (see Algorithm \ref{alg:alg_generation}). 
Note that the propagation graph can be further restricted to edges corresponding to a surrogate foreign key belonging to some potential key. 
Due to space constraints, this restriction of the propagation graph is omitted in this paper.

\subsubsection{Quantifying the artificial unicity}

We define a metric to quantify the level of artificial unicity in a relation.

\begin{definition}
Let $X$ be the potential key on $R$. The level of artificial unicity of $r$ with respect to a surrogate key $SK \in R \setminus X$ is defined as follows:
$$AU(r, X) = 1 - \frac{|[SK]_{X}^r|}{|r|}$$
\end{definition}

\subsubsection{Generation of an artificial unicity-free database}

 To get rid of artificial unicity, the two maps $clean\_SK$ and $clean\_FK$ are used while traversing the propagation graph in a certain order as shown in the following example. 

\begin{example}
The potential key of \texttt{Animal} is a foreign key pointing to a surrogate key in \texttt{Microchip} and the potential key of relation \texttt{Appointment} contains a foreign key pointing to a surrogate key in \texttt{Animal}, as presented in examples 1.2 and 3.2. 
Thus, artificial unicity in \textit{Perfect Pet}'s database should be removed one relation at a time, following the order: 1) \texttt{Microchip}, 2) \texttt{Animal}, 3) \texttt{Appointment}. 
\end{example}

The process of cleaning surrogate keys' values is described in Algorithm \ref{alg:alg_generation}.  
It is based on the oriented graph of surrogate keys-surrogate foreign keys associated to the database. 
A copy of the database is created on which key constraints are deactivated (lines 1-2). 
First, relations without outgoing edges, i.e. those relations without surrogate foreign keys are selected (line 5). Then, for each of them, the equivalence classes are extracted and the surrogate keys' values are cleaned (line 7). 
Afterwards, the equivalence is propagated to associated relations through Algorithm \ref{alg:fk_redundancy} (line 10). 
Finally the relation can be removed from $(V_{SK}, E_{SFK})$ (line 14).

\begin{example}
The version of the relation \texttt{Animal} free of artificial unicity obtained after application of Algorithm \ref{alg:alg_generation} to \textit{Perfect Pet}'s database is presented in table \ref{tab:cleaned_animal_rel}.
From this cleaned relation, $|\pi_{id\_microchip}(Animal')| = 3$, $|Animal'| = 7$, and $AU(\textsf{Animal'}, \\ \textsf{id\_microchip}) = \frac{3}{7}$.

\begin{table*}[bt]
	\caption{\texttt{Animal'}: artificial unicity free version of \texttt{Animal}}
	\centering
	\begin{scriptsize}
		\begin{tabular}{c|c|c|c|c|c|c|c|c|c|c}
			& \textbf{id\_animal} & \textbf{species} & \textbf{breed} & \textbf{name} & \textbf{id\_microchip} & \textbf{gender} & \textbf{dob}  & \textbf{weight} & \textbf{food} & \textbf{hash\_id}\\
			\hline
			1 & 744 & canine & greyhound & Gunter & 611 & M & 15-07-2014 & 27 & Greyh. Mix &  227f1df55c\\
			2 & 746 & canine & labrador & Coco & 613 & F & - & 41 & H Large Dog & 4745dd610e\\
			3 & 747 & feline & siamese & Izzy & 614 & F & 18-01-2018 & 4.2 & Indoor Cat & 9d5faf7fa6\\
			4 & 744 & canine & greyhound & Gunter & 611 & M & 15-07-2014 & 25.8 & Urinary & 227f1df55c\\
			5 & 747 & feline & siamese & Izzy & 614 & F & 01-01-2018 & 4.1 & Indoor Cat & 9d5faf7fa6\\
			6 & 746 & canine & labrador & Coco & 613 & F & - & 41 & H Large Dog & 4745dd610e\\
			7 & 744 & canine & greyhound & Gunter & 611 & M & 15-07-2014 & 26 & Urinary & 227f1df55c\\
			\hline
		\end{tabular}
	\end{scriptsize}
	\label{tab:cleaned_animal_rel}
\end{table*}
\end{example}

\begin{theorem}
Let $d' = Removing\_AU(d)$. \\
There does not exist any relation in $d'$ suffering from artificial unicity induced by surrogate keys in $d'$. 
\end{theorem}

We give a sketch of the proof by contradiction.
\begin{proof}
Let us assume that there exists a relation $r_1 \in d'$ such that $r_1$ suffers from artificial unicity induced by surrogate keys in $d'$. Let $SK_{r_1}$ be the surrogate key of $r_1$. 
Thus there exists two tuples $t_1, t_2 \in r_1$ such that $t_1[SK_{r_1}] \not= t_2[SK_{r_1}]$ and yet $t_1[SK_{r_1}] \equiv t_2[SK_{r_1}]$. 
\\
That means  that there exists a path in the propagation graph of $d'$, let us say $(r_1 \rightarrow r_2 \ldots \rightarrow r_{n-1} \rightarrow r_n)$  such that there exists two tuples $s_1, s_2 \in r_n$ with $s_1[SK_{r_n}] \equiv s_2[SK_{r_n}]$
and for every $ A \in X_{r_n}, s_1[A] = s_2[A]$
(see a similar illustration in figure \ref{fig:au_propag}), where 
 $SK_{r_n}$ (resp $X_{r_n}$) is the surrogate key (resp. potential key) of $r_n$.
 Note that the path cannot have a cycle since the propagation graph is assumed to be acyclic.
 \\
Since $r_n \in d'$ and $r_n$ is a leaf in the propagation graph, $r_n$ is equal to the result of  $clean\_SK(r, SK_r,X_r)$ for some $r\in d$.
We can deduce that $s_1[SK_{r_n}] = s_2[SK_{r_n}]$ .
\\
The same reasoning applies for every relation belonging to the inverse path from $r_n$ to $r_1$, leading at the end to $t_1[SK_{r_1}] = t_2[SK_{r_1}]$.
A contradiction.

\end{proof}

\begin{algorithm}
\caption{$Removing\_AU(d)$}
\label{alg:alg_generation}
\SetKwInOut{Input}{Input}
\SetKwInOut{Output}{Output}

\Input{A database $d$ over $\mathcal{R}$}
\Output{$d_{new}$ the artificial redundancy-free version of $d$}

$d_{new} = d$\;
Disallow all constraints on $d_{new}$ \;
Let $(V_{SK}, E_{SFK})$ be the propagation graph of $d$ \;
\While{$V \not= \emptyset$}{
        	$rel = \{ r \in V_{SK} \mid d^+(r) = 0 \}$\\
	\ForAll{$r \in rel$}{
		$r_{new} = clean\_SK(r, SK, X)$\;
		$d_{new} = (d_{new} \setminus \{r\}) \cup \{r_{new}\}$\;
		\ForAll{$s \in V_{SK}  \mbox{ such that } (s,r) \in E_{SFK} $}{
			$s_{new} = Remove\_FK\_Redundancy(s,r)$ \;
			$d_{new} = (d_{new} \setminus \{s\}) \cup \{s_{new}\}$ \;
			$E_{SFK}  = E_{SFK}  \setminus \{(s,r)\}$ \;
		}
	$V_{SK}  = V_{SK}  \setminus \{r\}$ \;
	}
}

\Return $d_{new}$\;
\end{algorithm}

\begin{algorithm}
\caption{$Remove\_FK\_Redundancy(s,r)$}
\label{alg:fk_redundancy}
\SetKwInOut{Input}{Input}
\SetKwInOut{Output}{Output}

\Input{$s, r$ two relations with at least one surrogate foreign key from $s$ to $r$}
\Output{$s_{new}$ the artificial redundancy-free version of $s$ with respect to $r$}

	\ForAll{surrogate foreign key from $FK$ of $s$ to $SK$ of $r$}{
		$s_{new} = clean\_FK(s, FK, r, SK)$\;
	}
\Return $s_{new}$\;
\end{algorithm}

\subsection{Normalisation and reduction}

Once artificial unicity is removed, redundancy and other data quality issues may still remain. 
This last step of RED2Hunt is devoted to cleaning the data for analytical purposes.

\subsubsection{Assessment of both data quality and normalization issues with respect to the natural key}

In a given relation where a natural key was identified, we are interested in qualifying how other attributes behave with respect to its natural key. To do so, we define the notion of \emph{attribute stability}, allowing the detection of both data quality and normalization issues. 
Intuitively, every functional dependency whose left hand side is the natural key and the right-hand side a remaining stable attribute should be satisfied. 
Any normalized relation should only include stable attributes with regards to its natural key. This is indeed not necessarily the case.

\begin{definition}
Let $r$ be a relation over $R$, $SK \in R$ a surrogate key, $X \subseteq R \setminus SK$ the natural key and $A \in R \setminus X \setminus \{SK\}$. \\
The level of stability of attribute $A$ with respect to $r$ and $X$ denoted by $Stability_{(r, X)}(A)$, is defined by:
$$Stability_{(r, X)}(A)= g_3(X \rightarrow A, r)$$
\end{definition}

A value of $Stability_{(r, X)}(A)$ close to zero, suggests that attribute $A$ is likely to belong to relation $r$. Conversely, the further this value is from zero, the more unstable the attribute seems to be, suggesting 
that it contributes to the denormalization of the relation and will eventually need to be removed or have its values cleaned. 
For the domain expert, the RRP associated to the relation provides useful insights: 
whenever the number of distinct values of some remaining attribute appears above the red line (i.e. is greater than that of $X$), it suggests some normalisation issues and instability of the attribute.
However, the opposite is not true, the number of unique values associated to attribute $A$ could be below the red line and value $Stability_{(r, X)}(A)$ could still be high, exhibiting its unstable character with regard to the natural key.

Remaining attributes that are neither surrogate keys nor potential keys are classified in two categories: 
\begin{itemize}
\item $AtbN$: contains all attributes considered unstable based on their value of $Stability$, and should probably be moved out of the relation or have their values cleaned.
\item $AtbC$: contains attributes considered stable, which should remain in the relation but whose values might require some cleaning.
\end{itemize}

The schema of any relation schema can be seen as follows: $R(SK, N, AtbN, AtbC)$.

\begin{example}
The values of $Stability$ in the relation \texttt{Animal'} give the following values:
 \texttt{species}: 0, \texttt{breed}: 0, \texttt{name}: 0, \texttt{gender}: 0, \texttt{dob}: $\frac{1}{7}$, \texttt{weight}: $\frac{3}{7}$, and \texttt{food}: $\frac{1}{7}$.
Therefore, $AtbN = \{\mathsf{dob}, \mathsf{weight}, \\ \mathsf{food}\}$ 
and 
$AtbC= \{\mathsf{species}, \mathsf{breed}, \mathsf{name}, \mathsf{gender}\}$. \\
Since  $AtbN$ is not empty, \texttt{Animal'} is likely to be denormalized. 
\end{example}

\subsubsection{Domain expert validation of attributes stability}

Data quality issues could bias the classification of attributes. One could appear to be unstable because of values or data format inconsistencies, or be classified as stable because of the low variation observed in its values at the time of the assessment. 
To validate or correct the stability category of the attributes, the domain expert is asked for each attribute whose $Stability$ measure's value is strictly above zero, if this attribute could potentially in real-life accept several values and still characterise the same entity. Here again data samples can be used to support her validation.

\begin{example}
The expert validates the classification of attributes \texttt{food} and \texttt{weight} in \texttt{Animal'} as unstable but corrects the classification of attributes \texttt{dob} in \texttt{Animal'} and \texttt{implant\_date} in \texttt{Microchip'} to stable. Indeed, an animal can only have one date of birth and a microchip can only be implanted once.
\end{example}

\subsubsection{Normalisation}

To fix the normalization problem in a given relation, we propose, as shown in Algorithm \ref{alg:alg_norm} to decompose the relation. The simple part (line 2-3) consists of only a projection on attributes not to be normalized ($AtbN$), and the other one addresses attributes of $AtbC$. The domain expert has to identify which attribute, or group of attributes, will complete the natural key of the original relation to become a natural identifier for unstable attributes' values (line 7). Quite often, this attribute $O$ is of type \texttt{date}, the new relation $r_2$ being an historisation of unstable attributes. Attribute $O$ could sometimes be part of another relation connected through some foreign keys.  
The validation has to be done by the domain expert (line 10).

\begin{algorithm}
\caption{$normalisation(r, d_{new})$}
\label{alg:alg_norm}
\SetKwInOut{Input}{Input}
\SetKwInOut{Output}{Output}

\Input{$d_{new}$ over $\mathcal{R}$, $r \in d_{new}$, $r$ over $R(SK, X, AtbC, AtbN)$ with $AtbN \not= \emptyset$}
\Output{$d_{new}$ the normalized new database}

$d_{new} = d_{new} \setminus \{r\}$ \;
$R_1 = \{SK\} \cup X \cup AtbC$\;
$r_1 = \pi_{R_1}(r)$ \;
$d_{new} = d_{new} \cup \{r_1\}$\;
finish = false\;
\While{not finish}{
	Let $O \in S, S \in \mathcal{R}$ such that $X \cup \{O\}$ is a natural key on $r_2$ \;
	Let $SK$ be a surrogate key of $r$\;
	$R_2 = \{SK\} \cup X \cup O \cup AtbN$ \;
	$r_2 = \pi_{R_2}(r)$\;
	\If{$\{SK, O\}$ is a key in $r_2$}
	{ 
		$d_{new} = d_{new} \cup \{r_2\}$\;
		finish = True\;
	}
}
\Return $d_{new}$\;
\end{algorithm}

\begin{example}
In \textit{Perfect Pet}'s example, \texttt{Animal'} is decomposed into two relations, one containing the stable attributes, and the other one the unstable attributes:\\
{
\footnotesize
\begin{flushleft}
\texttt{Animal'}(\underline{\texttt{id\_animal}}, \texttt{hash\_id\_animal}, \texttt{id\_microchip}, \texttt{species}, \texttt{breed}, \texttt{name}, \texttt{gender}, \textsf{dob}), \\
\texttt{Animal\_details}(\underline{\texttt{id\_animal}, \texttt{app\_date}}, \texttt{weight}, \texttt{food})
\end{flushleft}
}
Keys are underlined for the sake of clarity but still not declared in the new database at this stage.
The attribute selected by the expert to complete the natural key of the new relation is \texttt{Appointment.app\_date}. 
Relations \texttt{Microchip} and \texttt{Appointment} do not include any $AtbN$ attribute in their original form, thus do not need to be decomposed.                            
\end{example}

\subsubsection{Data cleaning}

After normalization, the schema of any relation can be seen as follows: $R(SK, X, AtbC)$. 
Attributes of $AtbC$ might still contain some inconsistent and inaccurate values. 
Algorithm \ref{alg:alg_chase} is applied to identify and resolve conflicting values of stable attributes with regards to the natural key, by applying the well-known \textit{chase} procedure to ensure that each FD of the form $X \rightarrow A, A \in AtbC$ is satisfied in its relation. Several conflict resolution methods are available, such as application of validation rules or training of a model to identify the accurate value. Their discussion is out of the scope of this paper. 
Clearly, the domain expert has to be implied whenever a counter example is identified (i.e. two tuples sharing the values of natural keys have conflicting values on stable attributes), to choose which one of the conflicting values has to be kept, unless she can identify with certainty a selection rule adapted to the attribute.

\begin{algorithm}
\caption{$chase(d)$}
\label{alg:alg_chase}
\SetKwInOut{Input}{Input}
\SetKwInOut{Output}{Output}

\Input{$d_{new}$}
\Output{$d_{clean}$, a clean version of $d_{new}$}

$d_{clean} = d_{new}$\;
\For{$r \in d_{clean}$}{
  let $r$ be over $R(SK, X, AtbC)$\;
  \If{$AtbC \not= \emptyset$}{
    \For{$A \in AtbC$}{
	$chase(r, X \rightarrow A)$ \;
     }
  }
}
\Return $d_{clean}$\;
\end{algorithm}

\subsubsection{Reduction}

At the end of the data cleaning step, we obtain a new normalized database that might still contain exact duplicates. To completely remove all the duplicates, each relation is replaced by the projection of its own tuples.

\begin{example}
\label{ex:sql_final}
In \texttt{Animal'}, two different \texttt{dob} were associated to \texttt{Izzy}. 
Application of \textit{chase} allowed the detection and correction of that error, before reduction of the relations. 
Cleaned relations \texttt{Animal'} and \texttt{Microchip'} are presented in table \ref{tab:animal_reduced}.

\begin{table*}
\caption{Extract of the redundancy-free database}
\label{tab:animal_reduced}
	\ContinuedFloat
		\begin{subtable}[b]{1\textwidth}
		\centering
			\begin{scriptsize}
			\begin{tabular}{c|c|c|c|c|c|c|c|c}
			& \textbf{id\_animal} & \textbf{species} & \textbf{breed} & \textbf{name} & \textbf{id\_microchip} & \textbf{gender} & \textbf{dob} & \textbf{hash\_id} \\
			\hline
			1 & 744 & canine & greyhound & Gunter & 611 & M & 15-07-2014 & 227f1df55c\\
			3 & 746 & canine & labrador & Coco & 613 & F & - & 4745dd610e\\
			4 & 747 & feline & siamese & Izzy & 614 & F & 18/01/2018 & 9d5faf7fa6\\
			\hline
			\end{tabular}
			\subcaption{relation animal'}
			\end{scriptsize}
		\end{subtable}

		\begin{subtable}[b]{1\textwidth}
		\centering
			\begin{scriptsize}
			\begin{tabular}{c|c|c|c|c}
			& \textbf{id\_animal} & \textbf{app\_date} & \textbf{weight} & \textbf{food} \\
			\hline
			1 & 744 & 08/02/2022 & 27 & Greyhound Mix \\
			3 & 746 & 20/04/2022 & 41 & H Large Dog \\
			4 & 747 & 10/05/2022 & 4.2 & Indoor Cat \\
			6 & 744 & 20/07/2022 & 25.8 & Urinary  \\
			7 & 747 & 20/07/2022 & 4.1 & Indoor Cat \\
			9 & 744 & 09/09/2022 & 26 & Urinary \\
			\hline
			\end{tabular}
			\subcaption{relation animal\_details}
			\end{scriptsize}
		\end{subtable}		
\end{table*}
\end{example}

Following this step, we have a new normalized redundant-free database, on which integrity constraints can be re-encoded.

\begin{example}
Continuing example \ref{ex:sql1}, the equivalent query below against the cleaned database returns now 3 tuples instead of 7 and indicates the accurate number of appointments each animal went to.
\begin{small}
\begin{verbatim}
SELECT id_microchip, name, species, breed, gender, 
COUNT(*) as num_appointment
FROM Animal an
JOIN Appointment ap ON an.id_animal = ap.id_animal
GROUP BY ap.id_animal
\end{verbatim}
\end{small}

\textit{Perfect Pet}'s newly generated database is ready to be used for any analytical use.

\end{example}


\section{IMPLEMENTATION AND EXPERIMENTS}

\subsection{Implementation}

The RED2Hunt framework was implemented as a python package on top of PostgreSQL, relying on several packages (pandas, psycopg2, matplotlib, networkx, among others). 
A flask web app was developed to facilitate its use, guiding the experts through the framework step by step, and offering the possibility to visualize data samples and counterexamples throughout the process. 

To address the data quality issues prevailing in real-life operational databases, several options were implemented in RED2Hunt such as values transformation and standardization. Matching of potential keys' values to define equivalence classes can be performed through the integration of any existing entity-matching method.

\subsection{Experiments}

To assess its scalability, RED2Hunt was applied to several synthetic databases including various levels of artificial unicity generated from the publicly available IMDB database \cite{imdb}. Synthetic databases are accessible at \cite{deteriorateddb}. 
RED2Hunt was also tested on and used to clean several operational relational databases with surrogate keys. 
None of these databases could be made public due to confidentiality concerns. 
Nevertheless, we present lessons learned from the method’s practical application, its execution, and the involvement of the domain expert.

\subsubsection{Scalability assessment}

\begin{figure}[h]
\caption{IMDB database schema comparison}
\begin{subfigure}[h]{0.4\textwidth}
	\centering
	\caption{Original}
	\label{fig:imdb_dm_original}
	\includegraphics[width=7cm]{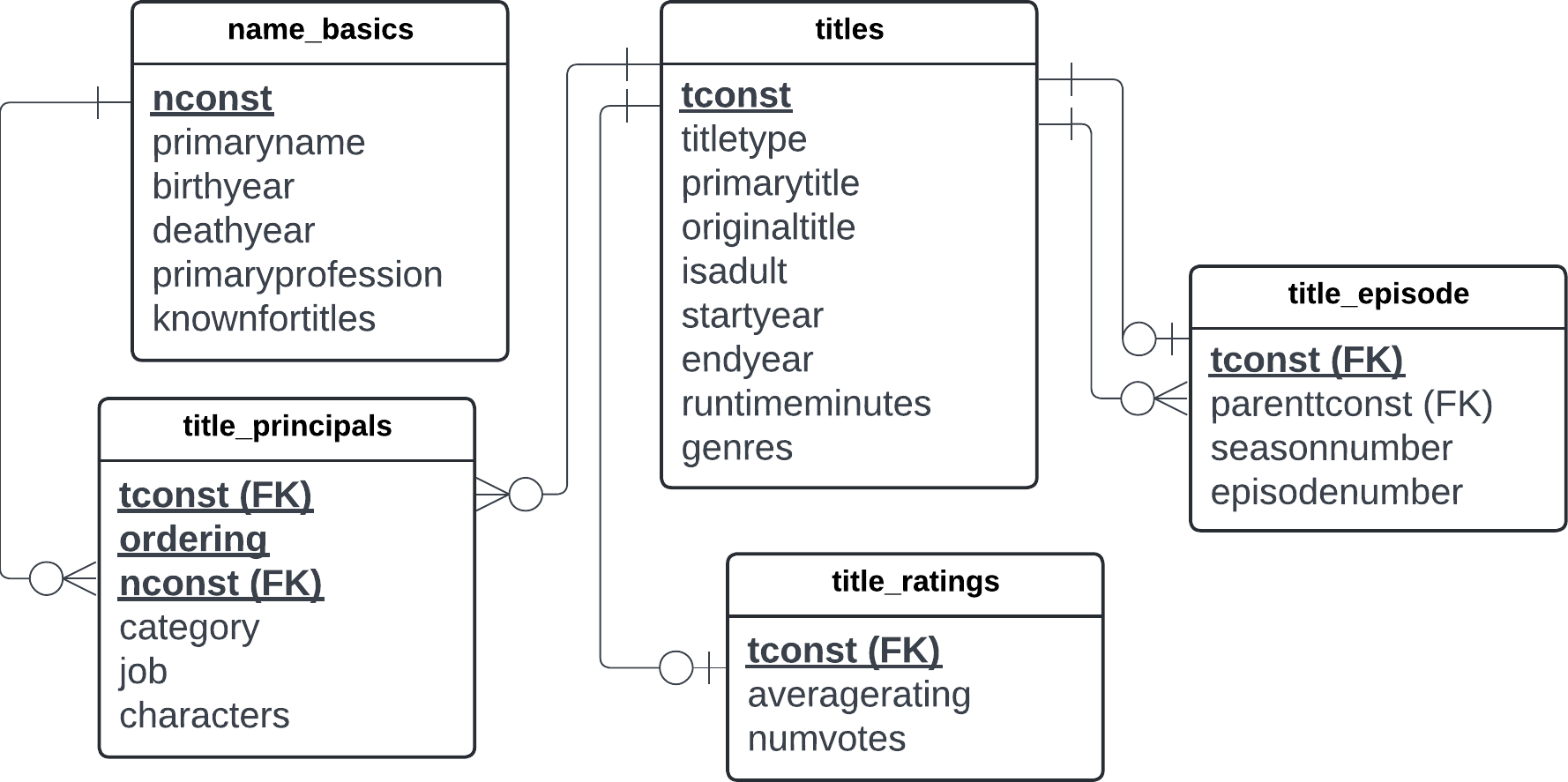}
\end{subfigure}
\hspace{0.1\textwidth}
\begin{subfigure}[h]{0.4\textwidth}
	\raggedleft
	\caption{Horizontally deteriorated}
	\label{fig:imdb_dm_deteriorated}
	\includegraphics[width=7cm]{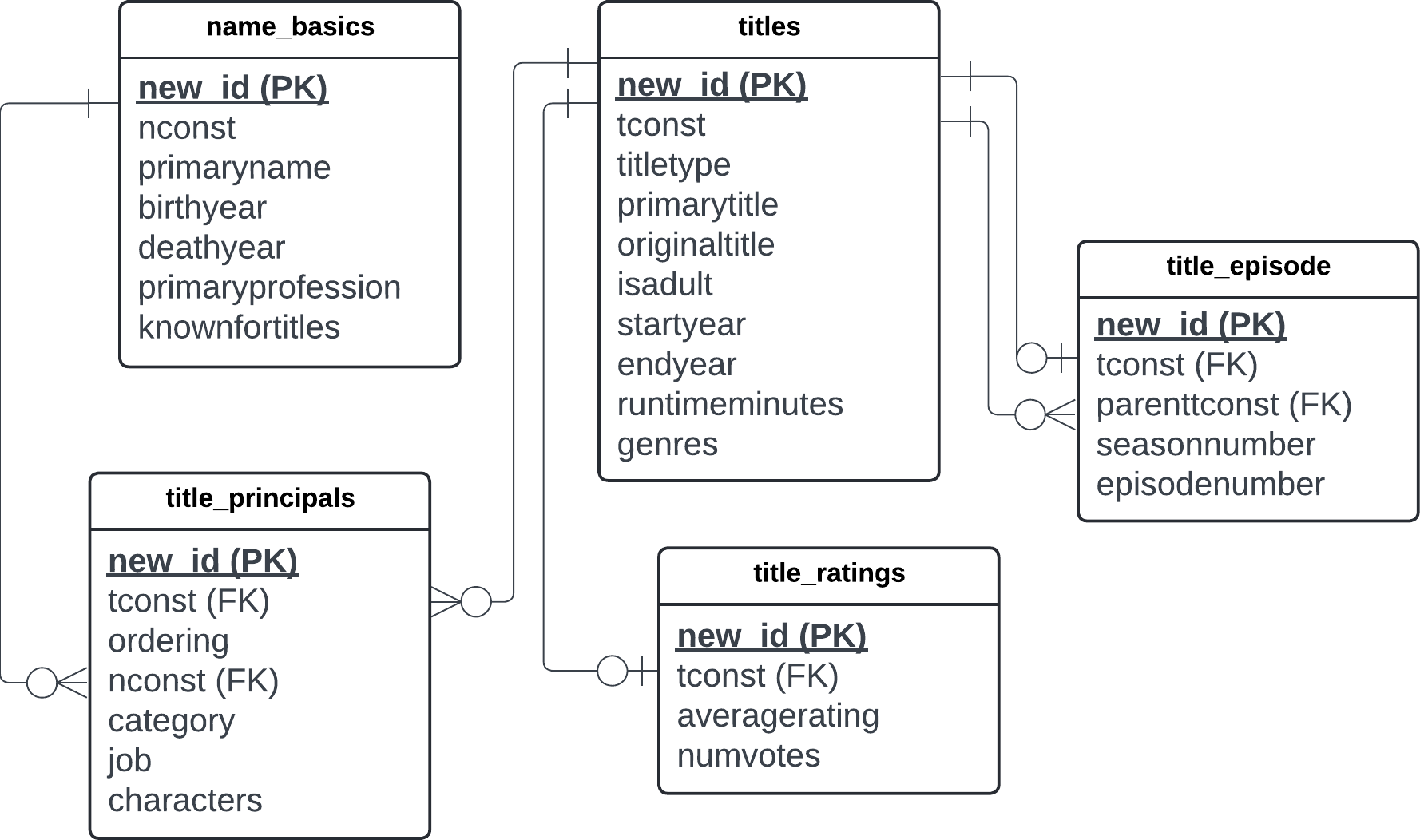}
\end{subfigure}
\end{figure}

\paragraph{Database deterioration} 
Copies of the IMDB database (see figure \ref{fig:imdb_dm_original}) were polluted to introduce artificial unicity, which is absent from its original version.
Seven deteriorated synthetic databases were generated by applying two deterioration patterns to introduce artificial unicity:
\begin{itemize} [leftmargin=*]
\item Tuple deterioration consists of injecting redundant tuples in every relation $r$ using a set deterioration factor $df_r$ (such that $|r| * df_r$ duplicates are injected) and a new surrogate primary key called $new\_id$ responsible for the artificial unicity in the relation, and modifying all foreign keys referencing the relation to point to this new surrogate key. Schema of databases suffering from tuple deterioration is presented in figure \ref{fig:imdb_dm_deteriorated}.
\item Attribute deterioration consists of duplicating non-natural key and non-surrogate key attributes in a relation on top of a tuple deterioration.
\end{itemize}

\begin{table}
\caption{Original IMDB's relations' sizes}
\begin{center}
\begin{scriptsize}
\centering
\begin{tabular}{c|c|c|c|c|c}
\textbf{\% AU} & \textbf{title} & \textbf{name\_basics} & \textbf{title\_episode} & \textbf{title\_ratings} & \textbf{title\_principals} \\ 
 \hline
original & 10,676,539 & 13,398,186 & 8,167,420 & 1,423,129 & 84,980,445\\
10\% & 11,744,193 & 14,738,005 & 8,984,162 & 1,565,442 & 93,478,489\\
20\% & 12,811,847 & 16,077,823 & 9,800,904 & 1,707,755 & 101,976,534\\
30\% & 13,879,501 & 17,507,642 & 10,617,646 & 1,850,068 & 110,474,578\\
40\% & 14,947,154 & 18,757,460 & 11,434,389 & 1,992,381 & 118,972,623\\
50\% & 16,014,808 & 20,397,279 & 12,251,130 & 2,134,698 & 127,470,667\\
\end{tabular}
\end{scriptsize}
\label{tab:imdb_size}
\end{center}
\end{table}

Assessment of RED2Hunt's scalability with regard to both the relations' sizes and relations' degrees is achieved by testing it on databases suffering from tuple and attribute deteriorations. 
For the experimentation’s purposes, we generated from the IMDB database five databases suffering from tuple deterioration using the deterioration factors 10\%, 20\%, 30\%, 40\%, and 50\%, and two databases suffering from attribute deterioration (on top of a tuple deterioration factor of 30\%) by doubling and tripling the non-SK, non-NK attributes. Table \ref{tab:imdb_size} presents the sizes of the relations in the databases suffering from tuple deterioration. \\

\emph{Method's application} 
This experiment’s goal was to qualify the computational performance of the method, not the expert’s involvement. 
We consider that in its original version, the IMDB database does not include any duplicate, thus the original database's primary keys were adopted as the set of natural keys (
\texttt{tconst} in \texttt{titles}, 
\texttt{nconst} in \texttt{name\_basics}, 
\texttt{tconst} in \texttt{title\_ratings}, 
\texttt{tconst} in \texttt{title\_episode}, 
and \texttt{tconst + nconst + ordering} in \texttt{title\_principals}) 
and only the newly added primary keys were considered as surrogate keys for the deteriorated databases. 
Furthermore, since the horizontal deterioration is achieved by an exact copy of the tuples, equivalence classes are drawn by exact match of the potential keys' values. 
The experimentations were conducted on a computer with 2.3 GHz Quad-Core Intel Core i7, and 32 GB RAM.  \\

\emph{Output comparison} 
The redundancy-free versions of the deteriorated databases obtained only differ from the original one on their relation schemas (because of the introduction of new primary keys and their associated foreign keys). This indicates that the framework allowed to find a database containing the same values as the original one, regardless of the deterioration factor and deterioration pattern. The number of tuples in every relation of every generated redundancy-free database matches exactly the size of the same relation in the original database.  All attribute values on deteriorated databases also match those of the original database. \\

\begin{figure*}[h]
\caption{Red2Hunt application execution times comparison on deteriorated databases}
\begin{subfigure}[h]{0.4\textwidth}
	\centering
	\caption{Total execution time by tuple deterioration factor}
	\label{fig:tot_exec_time_hor}
	\includegraphics[width=7.8cm]{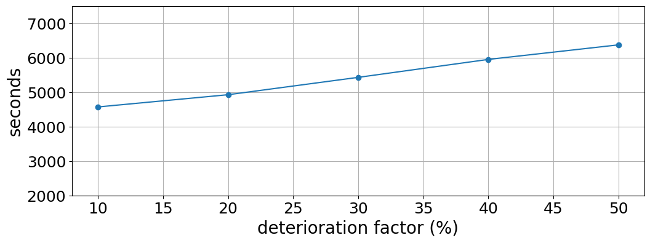}
\end{subfigure}
	\hspace{0.1\textwidth}
\begin{subfigure}[h]{0.4\textwidth}
	\centering
	\caption{Execution time breakdown by part of the framework and tuple deterioration factor}
	\label{fig:breakdown_exec_time_hor}
	\includegraphics[width=7.8cm]{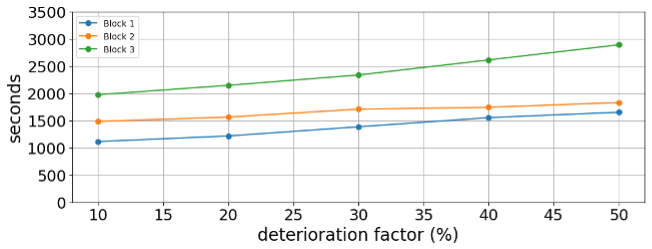}
\end{subfigure}

\begin{subfigure}[h]{0.4\textwidth}
	\centering
	\caption{Total execution time by attribute deterioration factor}
	\label{fig:tot_exec_time_vert}
	\includegraphics[width=7.8cm]{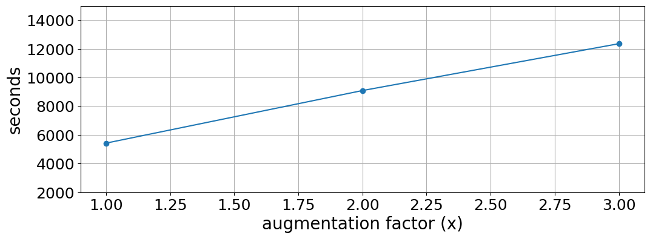}
\end{subfigure}
	\hspace{0.1\textwidth}
\begin{subfigure}[h]{0.4\textwidth}
	\centering
	\caption{Execution time breakdown by part of the framework and attribute deterioration factor}
	\label{fig:breakdown_exec_time_vert}
	\includegraphics[width=7.8cm]{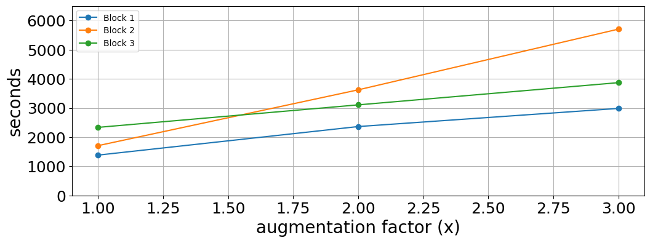}
\end{subfigure}
\end{figure*}

\emph{Execution time}
RED2Hunt was applied to each deteriorated database five times, and the average execution times for each block were calculated and are presented in figures \ref{fig:tot_exec_time_hor} and \ref{fig:breakdown_exec_time_hor} for the databases suffering from tuple deterioration only, and \ref{fig:tot_exec_time_vert} and \ref{fig:breakdown_exec_time_vert} for the ones suffering from both attribute and tuple deterioration. 
The execution time of the full framework's workflow is quasi-linear and proportional to the reduction factor (i.e. the size of the database). 
This particular trait is the strength of the framework compared to existing approaches relying on data-dependencies elicitation.
However, as observable on figures \ref{fig:breakdown_exec_time_hor} and \ref{fig:breakdown_exec_time_vert}, execution of block 2 is much more sensitive to the number of attributes than other parts, under the current implementation. The execution time of this part appears to be directly related to the augmentation factor.\\

\subsubsection{Application to operational databases}
RED2Hunt was applied to several relational databases used by SMEs and several valuable insights have emerged from these experiments, which are summarized below. 

First, two experts with different backgrounds are required to use RED2Hunt: a domain expert with good knowledge on the objects and events represented in the database, and a data expert to guide her through the process. 
Ideally, the domain expert should not be a database professional, and even less have participated in the design of the database, to avoid long (and tenuous) discussions on the data origin and modelling choices. \\

Second, with respect to surrogates keys occurring in practice, we noticed several things: first, the number of relations with at least one surrogate key were very high, up to 100\% in certain databases!  
Second, among those relations, a significant number (between 10 and 30\%) did not have any natural keys (declared or not), corresponding to potential keys that are not natural keys. For them, it was quite surprising to see that plenty of redundancy occurs for all attributes except the surrogate key, with reduction rates above 90\% for some relations with hundreds of thousands of tuples. 
For example we were able to reduce a relation containing originally 443,119 tuples to 20,075 "unique" tuples. \\

Third, we noticed that the order in which the relations are presented to the domain expert impacts the performance of Block 1. 
To allow her to familiarize seamlessly with the process we suggested organizing the interaction between three different "kinds" of relations: the first one corresponds to the leaves of the propagation graph. They are often easy to understand and allow the easing of the domain expert into the task. 
The second one corresponds to the relations with a limited number of attributes (less than 15) and a few foreign keys. 
Finally, the third one gathers the most complicated relations with possibly plenty of attributes and/or foreign keys. 
Another watch point was related to the presence of cycles in the propagation graphs. 
Indeed, we encountered some in the databases RED2Hunt was applied to. 
Restricting the graph to edges corresponding to surrogate foreign key belonging to some potential keys allowed us to address the problem and remove the artificial unicity within these databases, as all the restricted graphs were acyclic. 
The technical adaptations involved in this solution are not discussed in this paper. \\

Fourth, during the interaction, domain experts requested to examine data samples and counter-examples to validate or refuse hypotheses, including results of complex queries. 
The most important obstacle met was when discrepancy occurred between the real-world entities represented in a relation and the semantics conveyed by the name of the relation and its attributes. In these cases, the elicitation process was more complex and slower, and the domain expert had to resort to inspecting more data samples and counter-examples from the relation.
This phenomenon is often an indicator of denormalization, and our observations show that it is frequently related to the presence of temporal attributes and a relation allowing an historization of the attributes' values.  

We also noticed that RED2Hunt's concepts were easily and accurately grasped by the domain experts, who were able to quickly and precisely identify the natural and surrogate keys, data quality issues, and their causes. They indeed know their data the best. 

Another major observation made was that they were positively engaged throughout the entire process and willing to contribute more, which is very encouraging for concrete use of the framework. \\

Fifth, the execution times to run back-end computations (such as  RRP extraction, equivalence classes generation, and propagation, g3 calculation, etc.) were never seen as an obstacle.

The documented experience shows that experts spend between 3 and 12 minutes per relation to identify its natural and surrogate keys, depending on its position in the graph.
In a large database including 100 relations of interest for analytical use, this task would represent around two days of the expert’s time. 
This may appear substantial and confirms that there is "no free lunch": data quality issues cannot be resolved by miracle. 
Data analysts spend a significant amount of time cleaning their data for analytical or ML tasks, a process that is inherently challenging, if not impossible, to evaluate. 
We do believe that RED2Hunt brings something new in the landscape of data quality in databases with surrogate keys by tackling the problem at its root. \\

\section{Related work}

Numerous scientific topics are related to the content of our contribution, mainly data quality, duplicate detection, key constraint discovery, reverse engineering, 
and more broadly data exchange. \\

\emph{Data quality assessment and correction} approaches are usually based on rules-based validation \cite{fan2007improving}, statistical inference \cite{mayfield2010eracer} or machine learning \cite{mahdavi2019raha, liu2022picket}, and target specific data quality dimensions \cite{wang1996beyond}.
Holistic frameworks such as HoloDetect \cite{heidari2019holodetect} and HoloClean \cite{RekatsinasCIR17} have been developed to combine the advantages of different techniques and consider the collective impact of several data quality dimensions.  
Methods like ActiveClean \cite{krishnan2016activeclean} focus on cleaning data with regards to a specific analytical use. \\

\emph{Duplicate detection} also referred to as \textit{entity matching} or \textit{entity resolution}, is a subdomain of data cleaning, and is commonly applied to isolated datasets to identify pairs of duplicates among its observations representing the same individual entity in real-life. 
This can be achieved through different methods such as clustering, fuzzy matching, or even machine learning \cite{barlaug2021neural}. 
However, in relational databases, entity representations might be intricate. 
\cite{bhattacharya2007collective} use the associations between two entities of the same type to enhance entity resolution. \\

Data quality is usually validated and corrected by comparison to a "source of truth", which traditionally is domain expertise. Because it is considered a scarce resource, data cleaning frameworks based on alternatives have emerged such as knowledge base for KATARA \cite{Chi2015}, and crowd validation for CrowdER \cite{wang2012crowder} and Corleone \cite{gokhale2014corleone}. 
Unfortunately, industrial data, the target of RED2Hunt, often cannot be validated by comparison to these alternatives due to their private nature. 
ZeroER \cite{wu2020zeroer} performs entity resolution with unstructured learning and thus does not require any human involvement, but Li et al. \cite{li2021cleanml} demonstrated that its high-false positive rate could negatively impact the performance of a ML model. 
With the growing popularity of data quality as a research field, more studies are appearing about its research paradigm and evaluation in order to further improve outcomes \cite{sadiq2018data, panse2021evaluation}. \\

\emph{Key constraint discovery in relational databases}  \cite{bell1995discovery,huhtala1999tane,wyss2001fastfds} is an active and evolving field. 
This process is crucial to several data cleaning approaches such as rules-based validation and duplicate detection, and has been studied for years under different angles, from integrity constraints elicitation \cite{DBLP:journals/vldb/AbedjanGN15} including functional dependencies or keys \cite{bell1995discovery,huhtala1999tane,wyss2001fastfds,depuydt2018functional,DBLP:journals/jiis/JiangN20,DBLP:journals/tkde/CaruccioDNP21, wei2019discovery} or inclusion dependencies \cite{DBLP:journals/jiis/MarchiLP09}. 
Too many false positives are likely to be generated by inferring them from the database, due to noise and inconsistency of the data or the scalability in the number of attributes. 
We do believe that natural key discovery can be performed by domain experts with minimal human intervention using predefined rules and heuristics. 
Indeed, they are the only ones who really know their data. \\

\emph{Relational database reverse engineering} techniques focus on extracting high-level abstractions from existing relational databases. This process involves recovering the conceptual schema 
and understanding the underlying design, often for purposes like database migration or integration.
They provide insights into database design \cite{chiang1994framework, DBLP:journals/cacm/PremerlaniB94, DBLP:journals/ijcis/PetitTK95, hainaut1996database} by borrowing concepts from database theory, data mining, and software engineering. \\

\emph{Data exchange}
Data exchange in databases is crucial for enabling seamless data interoperability, integration, and migration across heterogeneous systems \cite{fagin2005data,lenzerini2002data,fagin2003data}. The research in this area combines database theory, algorithms, and practical tools to ensure that data can be accurately and efficiently transferred and transformed between different database systems.

\section{Conclusion}

In this article we defined the concept of \emph{artificial unicity}, a type of redundancy observable in operational databases with surrogate keys, preventing their straightforward cleaning for analytical purposes.
From our professional experience in dealing with operational databases, we argue that artificial unicity is not an isolated phenomenon and therefore requires special attention in the context of growing analytical practice within SMEs, especially considering the cost incurred by the exploitation of bad data.
To the best of our knowledge, this notion of artificial unicity induced by surrogate keys has not been identified so far by the database community.
\\

We proposed the RED2Hunt framework on top of PostgreSQL DBMS composed of three main steps: elicitation of keys, suppression of artificial unicity, and normalization and reduction, resulting in the generation of a redundancy-free database. Steps 1 and 3 require a domain expert who has to bring evidence about the data from her background knowledge.
The framework has been successfully tested on private SMEs' operational databases in the context of analytical projects, but because of their private nature, no such data could be made available for the community.
We used the publicly available database IMDB and produced many copies with synthetic artificial unicity and applied the framework on these data to assess its results and execution time. The outcome showed that the framework was able to generate databases without duplicates comparable to the original IMDB database, in quasi-linear times. The duration of the method's application was proportional to both the size of the database (as sum of its relations' sizes) and the number of attributes (as sum of its relation schemas' sizes). Not surprisingly, the proportionality coefficient is higher for the number of attributes than the number of tuples. 
\\

RED2Hunt is a comprehensive framework for cleaning dirty data due to the existence of surrogate keys, which addresses at the source several extremely difficult data quality problems. 
Its simplicity, scalability, and the quality of its outcomes should promote its adoption by SMEs willing to leverage their existing data. 
From our experience working on data-analytics projects and using RED2Hunt, we argue that the investment of time from an expert is lowered by using RED2Hunt on a medium-size database of 50 to 100 tables compared to the time required from her to manually clean a few datasets extracted from it. 
Moreover, the generated redundancy-free database simplifies the query process, as shown in example 4.20. 
We believe that this framework can improve decision making and resource management among companies adopting RED2Hunt. 
Such reductions open the door to financial and environmental impact improvement of the analytical activity.
\\

Currently, the treatment of relations in which natural keys have been declared is excluded from the RED2Hunt framework, as the elicitation of keys (Block 1) does not apply to them and they cannot suffer from any artificial unicity (Block 2). 
However, they could be candidate for decomposition and values conflict resolution (Block 3). 
As future work, we plan to extend  RED2Hunt to allow the treatment of such relations.

\bibliographystyle{IEEEtran} 
\bibliography{./biblio}

\end{document}